\documentclass[aps,pra,print,showpacs,superscriptaddress,twocolumn,
amsmath,amssymb]{revtex4-1}
\usepackage{amsfonts}
\usepackage{txfonts}
\usepackage{amsmath}
\usepackage{float}
\usepackage[colorlinks,breaklinks,linkcolor=blue,anchorcolor=blue,citecolor=blue,urlcolor=blue
]{hyperref}

\usepackage{graphicx}
\usepackage{dcolumn}
\usepackage{bm}
\usepackage{amssymb}
\usepackage{mathrsfs}
\usepackage[sort&compress]{natbib}
\usepackage{subfigure}

\begin{document}
\title{Counterdiabatic driving for pseudo- and antipseudo- Hermitian systems}
\author{Y. H. Song}
\affiliation{Center for Quantum Sciences and School of Physics, Northeast Normal University, Changchun 130024, China}
\author{Xin Wang}
\affiliation{Center for Quantum Sciences and School of Physics, Northeast Normal University, Changchun 130024, China}
\author{H. D. Liu}
\email[]{liuhd100@nenu.edu.cn}
\affiliation{Center for Quantum Sciences and School of Physics, Northeast Normal University, Changchun 130024, China}
\author{X. X. Yi}
\affiliation{Center for Quantum Sciences and School of Physics, Northeast Normal University, Changchun 130024, China}
\date{\today}

\begin{abstract}
In this work, we study the counterdiabatic driving scheme in pseudo- and antipseudo- Hermitian systems. By discussing the adiabatic condition for non-Hermitian system, we show that the adiabatic evolution of  state can only be realized in the non-Hermitian system which possesses real energy spectrum. Therefore, the counterdiabatic driving scheme to reproduce an exact evolution of an energy eigenstate needs either real energy spectrum or dropping its parts of dynamic phase and Berry phase. In this sense, we derive the adiabatic conditions and  counterdiabatic driving Hamiltonians for the pseudo-Hermitian Hamiltonian which possesses either real or complex energy spectrum and  the antipseudo-Hermitian Hamiltonian which possesses either imaginary or complex energy spectrum. We also find the condition to get self-normalized energy eigenstates in pseudo- and antipseudo- Hermitian system and derive the well-defined population of bare states on this energy eigenstate. Our results are illustrated by studying the counterdiabatic driving for a non-Hermitian three level system, and a perfect population transfer with loss or gain is realized.
\end{abstract}

\maketitle

\section{Introduction}
Adiabatic evolution \cite{Born1928} and geometric phase \cite{Berry1984} are two important concepts which play essential roles in quantum information process and quantum computation on both theoretical and experimental aspect. However, the long evolution time required by the adiabatic condition makes the quantum adiabatic process easily affected by environmental noise and decoherence \cite{RevModPhys.70.1003,RevModPhys.79.53,PhysRevA.96.042104,PhysRevA.102.023515,PhysRevE.88.062122,Aharonov1987}. To address this issue, the technology known as  shortcuts to adiabaticity (STA) \cite{PhysRevLett.104.063002,Guery-Odelin2019}, which can ``accelerate" adiabatic processes, has been developed in recent years \cite{Berry2009,Muga2009,PhysRevA.82.053403,Chen2011,PhysRevA.83.043804,PhysRevA.84.031606,PhysRevA.83.013415,PhysRevLett.105.123003,PhysRevA.89.012326,PhysRevA.89.033856,PhysRevA.89.063412,PhysRevA.91.053406,PhysRevA.94.063411}. Compared with the adiabatic process, the STA process can produce the same results (such as populations, states and Berry phases \cite{Berry1984,Berry2009}) as the slow, adiabatic process in a limited and shorter time. Consequently, this technology \cite{PhysRevLett.104.063002,PhysRevLett.124.150603,PhysRevResearch.2.023328} has been expanded and applied into atomic \cite{PhysRevLett.105.123003,Pancharatnam1956,Rigolin2014}, molecular and optical physics, such as the rapid transmission of ions and neutral atoms, the manipulation of internal populations, the preparation of states, and the expansion of cold atoms \cite{PhysRevA.100.012111,PhysRevE.98.032136}.

Among the STA schemes, the superadiabatic quantum driving \cite{Demirplak2003,Demirplak2005,Demirplak2008,Masuda2015} proposed by Demirplak and Rice, also known as counterdiabatic (CD) driving \cite{Berry2009,PhysRevX.4.021013,Jain1998,Gonzalez2007,Bruno2004}, is a particularly influential one. It uses an external filed or interaction to eliminate non-adiabatic coupling for a time-dependent Hamiltonian $H_0(t)$ to ensure the new Hamiltonian $H(t)$ has a same Schr\"{o}dinger equation solution as that under the adiabatic approximation. The new Hamiltonian constructed by CD driving can drive the system to evolve exactly along the adiabatic path of $H_0(t)$ \cite{PhysRevA.93.052109,Higgins1958,Rigolin2012,Zhang2016,
PhysRevA.89.012123,*PhysRevA.98.022102,*PhysRevA.93.043419,*dou2017high,*zhang2021high}.

However, if the environmental noise and decoherence are taken into consideration, the original Hermitian problem will become a non-Hermitian problem in an open system. In recent years, with the rapid progress of experiments in non-Hermitian physics in these photonics, phononics, condensed matter and cold atom systems,  non-Hermitian physics has drawn much more attention\cite{El-Ganainy2018,Ashida2020,Miri2019,Zhu2018,Wu2019,Li2019}.
In recent years, the non-Hermitian physics has drawn much more attention. In photonics, phononics and circuit systems, by introducing properties such as gain, loss and non-reciprocal coupling, their dynamics are characterized by non-Hermitian effective Hamiltonians\cite{El-Ganainy2018,Ashida2020,Miri2019}. In addition, the non-Hermitian properties of quantum systems have also been observed in experiment by using measurement and postselection \cite{Zhu2018,Wu2019,Li2019}. Because of its richer structure than Hermitian quantum systems, non-Hermitian systems have become a new platform for exploring novel quantum states and physical phenomena \cite{PhysRevA.99.032121,PhysRevLett.89.270401,Bender_2007,Berry_2011,PhysRevLett.103.123601,PhysRevLett.104.054102,Bender_2003,PhysRevLett.120.146402}.  Especially, some non-Hermitian Hamiltonians can also have pure real energy spectra, e.g. those with parity-time (PT) symmetry \cite{PhysRevLett.80.5243} or pseudo Hermiticity \cite{Mostafazadeh2002,*Mostafazadeh2002a,*Mostafazadeh2002b}, their exceptional points (EPs) of the energy spectrum in the parameter space are closely related to the symmetry, topological properties, and phase transitions of the system \cite{El-Ganainy2018,Ashida2020,Miri2019}.

Consequently, the Berry phase, adiabatic evolution and its shortcuts in non-Hermitian system  become a new focus in Hermitian physics.  Many efforts have been devoted into these topics, such as population inversion by CD driving \cite{PhysRevA.84.023415,*PhysRevA.86.019901}, using non-Hermiticity cancel the non-adiabatic coupling \cite{Torosov2013,Torosov2014} and designing new quantum annealing  algorithm \cite{Nesterov2012}.
However, the adiabatic condition \cite{Sun1993,PhysRevA.89.033403} and Berry phases \cite{Sun1993,Dattoli1990,PhysRevA.99.032121} in non-Hermitian systems are quite different with those in Hermitian systems. The CD driving schemes in non-Hermitian systems has been only developed for weak non-Hermitian Hamiltonian \cite{PhysRevA.84.023415,*PhysRevA.86.019901,PhysRevLett.109.100403,PhysRevA.93.052109,Song2016,Lig2017,Li2017} or used gain and loss as resource of CD driving \cite{Torosov2013,Torosov2014,Wu2016,Lig2017}.  Hence, we are inspired to ask what kind of non-Hermitian Hamiltonian can evolve adiabatically? How does one accelerate the adiabatic evolution of a non-Hermitian Hamiltonian by CD driving?  Furthermore, the populations of the bare states on the adiabatic eigenstates can not be normalized by a time independent coefficient \cite{PhysRevA.89.033403}. This makes it difficult to realize the applications of shortcut for adiabaticity, e.g., population inversion or transfer, in non-Hermitian system.  Therefore, it is naturally to ask do the left and right eigenstates for the non-Hermitian Hamiltonian can be the same one? Can the CD driving of a self-normalized eigenstate be realized by a non-Hermitian driving Hamiltonian? In this work, we attempt to shed light on these questions in  pseudo-Hermitian system \cite{Mostafazadeh2002,*Mostafazadeh2002a,*Mostafazadeh2002b} whose Hamiltonian has real or paired complex conjugate eigenvalues and antipseudo-Hermitian system whose energy eigenvalues are imaginary or have opposite real parts. Their left and right eigenstates can be related by a unitary transformation. According to the adiabatic condition \cite{Sun1993,PhysRevA.89.033403},  real energy spectrum is normally needed for the non-Hermitian Hamiltonian evolving adiabatically. For this reason, we study the adiabatic evolution, Berry phase and CD driving scheme in these two kinds of non-Hermitian system. Besides, we also discuss the condition of a non-Hermitian Hamiltonian possessing a self-normalized eigenstate.


This paper is organized as follows. In Sec. \ref{secii}, we first introduce the adiabatic evolution, geometric phase, 
and the CD driving for the Hermitian and non-Hermitian system. Then, the CD driving for the pseudo-Hermitian and antipseudo-Hermitian system are discussed in Sec. \ref{seciv}. In Sec. \ref{secv}, a model of non-Hermitian three-level system whose Hamiltonian can be pseudo-Hamiltonian or antipseudo-Hermitian is studied to illustrate our theory, we discuss its Berry connection, adiabatic evolution process, and calculate the adiabatic shortcut. And the corresponding diagrams are made for comparison and analysis. Finally, we conclude our results in Sec. \ref{secvi}.
\section{CD driving for non-Hermitian system  \label{secii}}
\subsection{CD driving for Hermitian system}
 We first introduce the CD driving of Hermitian system. Consider an arbitrary time-dependent Hamiltonian ${H}_{0}(t)$, with instantaneous eigenstates and energies given by
\begin{equation}
{H}_{0}(t)\left|E_{n}(t)\right\rangle=E_{n}(t)\left|E_{n}(t)\right\rangle.
\end{equation}
If its evolution satisfies the adiabatic condition
\begin{equation}
\frac{|\hbar\langle E_{m}(t) \mid \dot E_{n}(t)\rangle|}{|E_m-E_n|} \ll 1,
\end{equation}
the time dependent state
\begin{equation}
\begin{split}
\left|\psi_{n}(t)\right\rangle=&\exp \left\{-\frac{i}{\hbar} \int_{0}^{t} d t^{\prime} E_{n}\left(t^{\prime}\right)\right.\\
&\left.-\int_{0}^{t} d t^{\prime}\left\langle E_{n}(t)\left(t^{\prime}\right) | \partial_{t^{\prime}}E_{n}(t)\left(t^{\prime}\right)\right\rangle\right\}|E_{n}(t)\rangle
\end{split}
\end{equation}
which is initially an eigenstate $|\psi_n(0)\rangle=|E_n(0)\rangle$, will still be an eigenstate for $E_n(t)$ with dynamic phase $-\frac{1}{\hbar} \int_{0}^{t} d t^{\prime} E_{n}(t^{\prime})$ and geometric phase $i\int_{0}^{t} d t^{\prime}\left\langle E_{n}(t)\left(t^{\prime}\right) | \partial_{t^{\prime}}E_{n}(t)\left(t^{\prime}\right)\right\rangle$. By the reverse engineering approach \cite{Berry2009,PhysRevA.82.053403}, this exact adiabatic evolution can be reproduced by a new Hamiltonian ${H}(t)$
\begin{equation}
\begin{split}
{H}(t)=&\sum_{n}|E_{n}(t)\rangle E_{n}\langle E_{n}( t )|+i \hbar
\sum_{n}\left(
\left|
\partial_{t} E_{n}( t )\rangle\langle E_{n}( t )
\right| \right.\\
&\left.-\langle E_{n}( t ) | \partial_{t} E_{n}( t )\rangle
\left|E_{n}( t )\rangle \langle E_{n}( t )\right|
\right)\\
\equiv&{H}_{0}(t)+{H}_{1}(t)
\end{split}
\end{equation}without the adiabatic approximation, by adding an extra part
\begin{equation}
{H}_1(t)=i \hbar
\sum_{n}\left(
\left|
\partial_{t} E_{n}( t )\rangle\langle E_{n}( t )
\right|-\langle E_{n}( t ) | \partial_{t} E_{n}( t )\rangle
\left|E_{n}( t )\rangle \langle E_{n}( t )\right|
\right)
\end{equation}
which eliminates the contribution of energy crossing and generating the adiabatic geometric phase.
\subsection{Biorthonormal bases for non-Hermitian Hamiltonian}
The eigen problem and dynamics of the non-Hermitian are quite different with the Hermitian one. Due to the non-Hermicity, the eigenbasis of a non-Hermitian time-dependent Hamiltonian $H$  becomes a pair of biorthonormal eigenbasis. The right eigenstate  $\left|E_{n}^{r}(t)\right\rangle$ and the left eigenstate $\left|E_{n}^{l}(t)\right\rangle$ satisfy the energy eigen equations
\begin{equation}
{H}(t)\left|E_{n}^{r}(t)\right\rangle=E_{n}(t)\left|E_{n}^{r}(t)\right\rangle
\label{eq:1}
\end{equation}
and\begin{equation}
{H}^{\dagger}(t)\left|E_{n}^{l}(t)\right\rangle=E_{n}^{*}(t)\left|E_{n}^{l}(t)\right\rangle,
\label{eq:2}
\end{equation}
respectively. With both the right and left eigenstates, one can establish the orthonormal relation and the closure relations
\begin{equation}
\begin{aligned}
\left\langle E_{m}^{l}(t) \mid E_{n}^{r}(t)\right\rangle&=\left\langle E_{m}^{r}(t) \mid E_{n}^{l}(t)\right\rangle=\delta_{m n},\\
\sum_{n}  |E_{n}^{r}(t)\rangle  \langle E_{n}^{l}(t)|&=\sum_{n}| E_{n}^{l}(t)\rangle\langle E_{n}^{r}(t)|=1.
\end{aligned}
\end{equation}
This biorthonormal relation means that we need both the left and right eigenbasis to represent a generic state $|\psi\rangle$ as $|\psi\rangle=\sum_nC_n|E_n^r(t)\rangle$ with $C_n=\langle E_n^l|\psi\rangle$. Unlike the Hermitian case, the dynamic evolution of $|\psi\rangle$ is normally non-unitary. However, the non-unitary evolution under non-Hermitian can also be interpreted as a normalized state with a real phase factor like in the Hermitian system and a pure imaginary phase which corresponds gain or loss introduced by the non-Hermitian part  by the projective Hilbert space method (see Appendix \ref{PHS}, its further applications will be discussed in the future). Next, we will discuss the adiabatic evolution and shortcut to adiabaticity in the non-Hermitian system.
\subsection{Adiabatic evolution and shortcut to adiabaticity for non-Hermitian system}
To discuss the adiabatic evolution in non-Hermitian system, we suppose a time-dependent generic state can be decomposed by the right eigenbasis as
\begin{equation}
|\psi(t)\rangle=\sum_{n} c_{n}(t) \exp \left[-\frac{i}{\hbar} \int_{0}^{t} E_{n}\left(t^{\prime}\right) d t^{\prime}\right]\left| E_{n}^{r}(t)\right\rangle.
\label{eq:jr}
\end{equation}
Unlike the Hermitian system, the adiabaticity condition for non-Hermitian system takes the form \cite{Sun1993,PhysRevA.89.033403}
\begin{equation}
\frac{|\langle E_{n}^{l}(t) \mid \dot{E}_{m}^{r}(t)\rangle|}{\left|\omega_{n m}(t)\right|} e^{-\operatorname{Im}\left[W_{n m}(t)\right]} \ll 1,
\label{eq:jrjs}
\end{equation}
where $\omega_{n m}(t)=\left[E_{n}(t)-E_{m}(t)\right] / \hbar$ and $W_{n m}(t)=\int_{0}^{t} \omega_{n m}\left(t\right) d t$.
Under this adiabatic condition, if the initial state is a right eigenstate $|\psi(0)\rangle=|\psi^r_n(0)\rangle$, the state
\begin{equation}
\begin{split}
\left|\psi_{n}^{r}(t)\right\rangle=&\exp \left\{-\frac{i}{\hbar} \int_{0}^{t} d t^{\prime} E_{n}\left(t^{\prime}\right)\right.\\
&\left.-\int_{0}^{t} d t^{\prime}\left\langle E_{n}^{l}\left(t^{\prime}\right) | \partial_{t^{\prime}}E_{n}^{r}\left(t^{\prime}\right)\right\rangle\right\}|E_{n}^{r}( t )\rangle
\end{split}
\label{xde}
\end{equation}
will still be the time-dependent right eigenstate, with non-Hermitian adiabatic geometric phase

\begin{equation}
\gamma_{n}=\int_0^t A_n(t') dt',
\end{equation}
where
\begin{equation}
A_n(t)\equiv i\left\langle E_{n}^{l} (t)\left|\frac{\partial}{\partial t}\right|E_{n}^{r} (t)\right\rangle.
\end{equation}
is the Berry connection.

Similar with the Hermitian system, the CD driving Hamiltonian for this adiabatic evolution can be constructed by \cite{PhysRevA.84.023415}
\begin{equation}
\begin{split}
{H}(t)&={H}_{0}(t)+{H}_{1}(t)\\
H_0&\equiv\sum_{n}|E_{n}^{r}( t )\rangle E_{n}\langle E_{n}^{l}( t )|\\
H_1&\equiv i \hbar
\sum_{n}\left(
\left|
\partial_{t} E_{n}^{r}( t )\rangle\langle E_{n}^{l}( t )
\right| -\langle E_{n}^{l}( t ) | \partial_{t} E_{n}^{r}( t )\rangle
\left|E_{n}^{r}( t )\rangle \langle E_{n}^{l}( t )\right|
\right)
\end{split}
\end{equation}
However, not all the evolution from $|E_n^r(0)\rangle$ to $|\psi^r_n(t)\rangle$ are adiabatic. Notice the exponential part of Eq. (\ref{eq:jrjs}), the imaginary part of energy level  spacing will always cause transition between energy levels no matter how small the term $\langle E_n^l(t)|\dot{E}_m^r(t)\rangle$ is. So, there are two ways to satisfy this adiabatic condition: $\langle E_n^l(t)|\dot{E}_m^r(t)\rangle=0$ or $\operatorname{Im}\left[W_{n m}(t)\right]=0$. Therefore, the non-Hermitian Hamiltonian normally should possess full real spectrum to realize adiabatic evolution. Consider the pseudo-Hermitian system is one type of non-Hermitian systems which can possess full real spectrum, we next study how to accelerate the adiabatic evolution of pseudo-Hermitian system by CD driving.
\section{ Shortcut to adiabaticity for pseudo-and antipseudo- Hermitian system\label{seciv}}
\subsection{pseudo-Hermitian system}
For a pseudo-Hermitian system, its Hamiltonian satisfies \cite{Mostafazadeh2002}
\begin{equation}
{H}_p^{\dagger}={U} {H}_p {U}^{-1},
\label{eq:dia_1}
\end{equation}
where  ${U}$ is the symmetry matrix which is a unitary and Hermitian operator. The reason why the pseudo-Hermitian Hamiltonian can have real energy spectrum is that it has the same secular equation
\begin{equation}
\begin{aligned}
\det({H}_p^{\dagger}-E{I})&=\sum_na_n^*E^{n}\\
&=\det({U} {H}_p {U}^{-1}-E{I})\\
&=\det({H}_p-E{I})=\sum_na_nE^{n}=0
\end{aligned}
\end{equation}
with its Hermite conjugate $H^\dagger_p$. Therefore, the coefficients of the equation are all real: $a_n=a_n^{*}$, and the eigenenergies of $H_p$ are either real or paired complex conjugate with each other, i.e., if $E_n$ is the one eigenvalue of $H_p$,  $E_n^*$ must be another eigenvalue. $H_p$ and $H_p^\dagger$ share the same energy spectrum. By the time dependent eigen equation
\begin{equation}
{H}_p(t)\left|E_{n}^{*r}(t)\right\rangle=E_{n}^{*}(t)\left|E_{n}^{*r}(t)\right\rangle
\label{eq:right-1}
\end{equation}
for right eigenstate $\left|E_{n}^{*r}(t)\right\rangle$ and notice the eigen equation
\begin{equation}
{U}^\dagger(t){H}_p^{\dagger}(t)\left|E_{n}^{l}(t)\right\rangle={H_p}(t){U}^{\dagger}(t)\left|E_{n}^{l}(t)\right\rangle=E_{n}^{*}(t)U^\dagger(t)\left|E_{n}^{l}(t)\right\rangle
\label{eq:left}
\end{equation}
for left eigenstate $\left|E_{n}^{l}(t)\right\rangle$, we can easily get that
\begin{equation}
\left|E_{n}^{l}(t)\right\rangle={U}_n(t)\left|E_{n}^{*r}(t)\right\rangle
\label{fl}
\end{equation}
\begin{figure}[t]
    \centering
    \includegraphics[width=\columnwidth]{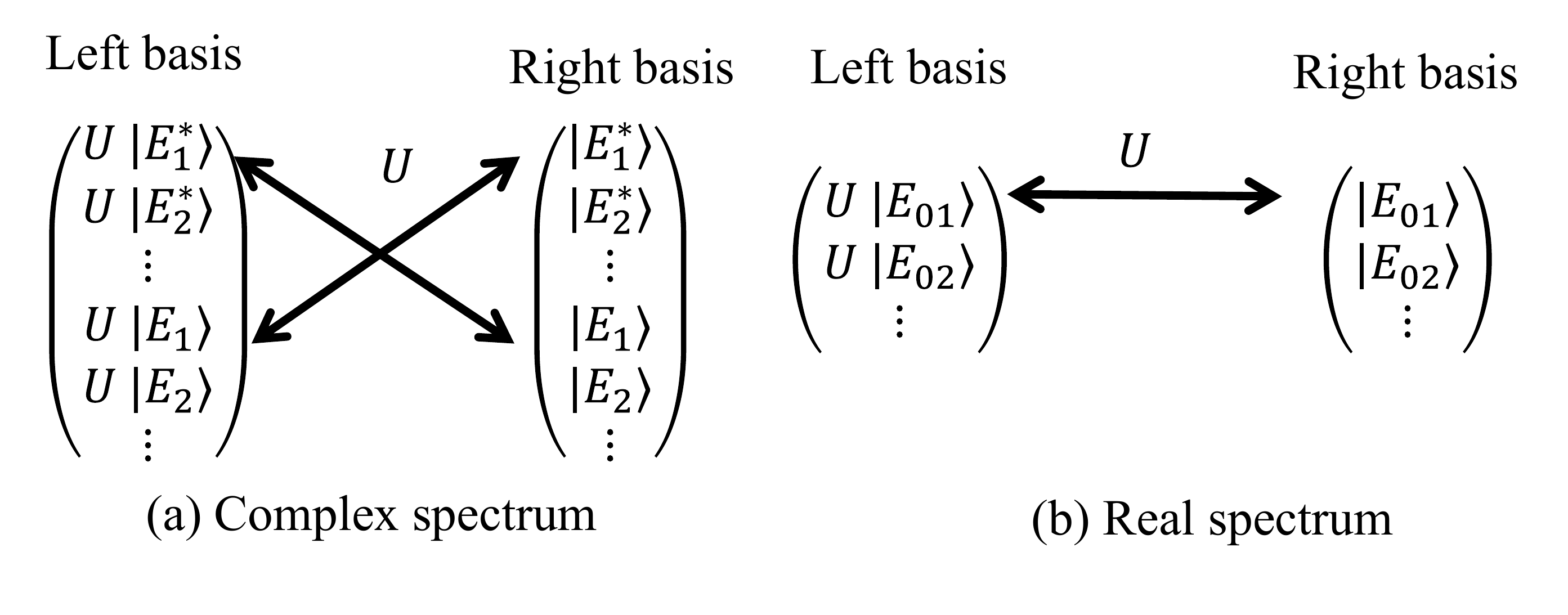}
\caption{Schematic illustration of the connection between the left and right basis in pseudo-Hermitian system}
    \label{fig1}
\end{figure}
for a non-degenerate energy spectrum, where ${U}_n(t)\equiv {U}(t)e^{i\phi_n}$ and $\phi_n$ is a constant phase factor. This means that the left and right eigenbasis can be connected by a unitary transformation $U(t)$ as shown in Fig. \ref{fig1}.  Therefore, we only need one set of right states $|E_n^r(t)\rangle\equiv |E_n(t)\rangle$ to study the non-Hermitian system. For the complex spectrum, the eigenstate $|E_n(t)\rangle$ of $E_n(t)$ is connected with the eigenstate $|E^*_n(t)\rangle$ of its complex conjugate $E_n^*(t)$. The orthonormal relationships between the left and right eigenbasis then become
\begin{equation}
\langle E_{m}(t)|{U}_n(t)| E_{n}^{*}(t)\rangle=\langle E_{m}^{*}(t)|{U}_n(t)|E_{n}(t)\rangle=\delta_{m n}.
\label{eq:8}
\end{equation}
Therefore, the adiabatic condition (\ref{eq:jrjs}) can be written as
\begin{equation}
\frac{|\left\langle E_{n}^{*}(t)\right|{U}(t)\mid \dot{E}_{m}(t)\rangle|}{\left|\omega_{n m}(t)\right|} e^{-\operatorname{Im}\left[W_{n m}(t)\right]} \ll 1.
\label{pscon1}
\end{equation}
Under this adiabatic condition, the Berry connection
\begin{equation}
A _{n}(t)=i\left\langle E_{n}^{*}( t )\left|{U}_{n}(t) \right|\dot{E}_{n}( t )\right\rangle
\label{eq:berrysp1}
\end{equation}
is decided by the pairs of eigenstates for the eigenvalues $E_{n}( t )$ and $E_{n}^{*}( t )$ .

For a full real spectrum, the orthonormal relation simply becomes  $\langle E_m(t)|U_n(t)|E_n(t)\rangle=\delta_{mn}.$ The adiabatic condition and Berry connection become
\begin{equation}
\left|\frac{\langle E_{n}^{r}(t)|{U}(t)| \dot{E}_{m}^{r}(t)\rangle}{\omega_{n m}(t)}\right| \ll 1
\label{pscon}
\end{equation}
and
\begin{equation}
A _{n}(t)=i\left\langle E_{n}( t )\left|{U}_{n}(t) \frac{\partial}{\partial t }\right| E_{n}( t )\right\rangle,
\label{eq:berryspr}
\end{equation}
which are similar to the Hermitian ones. It is worth to note that, $A_n(t)$ is usually not real if $U(t)$ is time dependent. Even the adiabatic condition (\ref{pscon}) is satisfied, the imaginary part of $A(t)$ can still destruct the adiabatic evolution.

Next, we introduce the CD driving for the pseudo-Hermitian system. Consider a time-dependent pseudo-Hermitian Hamiltonian ${H_0(t)}$, the  instantaneous eigenstate and energy are given by
\begin{equation}
{H}_{0}(t)|E_{n}( t )\rangle=E_{n}(t)|E_{n}( t )\rangle.
\end{equation}
Under the adiabatic condition (\ref{eq:jrjs}), if $\psi_n(0)=|E_n(0)\rangle$, the evolution of $|\psi_n(t)\rangle$ is described by
\begin{equation}
\begin{split}
\left|\psi_{n}(t)\right\rangle=&\exp \left\{-\frac{i}{\hbar} \int_{0}^{t} d t^{\prime} E_{n}\left(t^{\prime}\right)\right.\\
&\left.-\int_{0}^{t} d t^{\prime}\left\langle E_{n}^{*}\left(t^{\prime}\right)\left|{U}_n(t')\right| \partial_{t^{\prime}}E_{n}\left(t^{\prime}\right)\right\rangle\right\}|E_{n}( t )\rangle.
\end{split}
\end{equation}
This adiabatic evolution can be realized by a non-unitary operator
\begin{equation}
\begin{split}
{V}(t)=&\sum_{n} \exp \left\{-\frac{i}{\hbar} \int_{0}^{t} d t^{\prime} E_{n}\left(t^{\prime}\right)\right.\\
&\left.-\int_{0}^{t} d t^{\prime}\left\langle  E_{n}^{*}\left(t^{\prime}\right)\left|{U}_n(t')\right| \partial_{t^{\prime}}E_{n}\left(t^{\prime}\right)\right\rangle\right\}|E_{n}( t )\rangle\langle E^*_{n}( 0 )|U(0)
\end{split}
\end{equation}
which satisfies ${V}(t)\left|\psi_{n}(0)\right\rangle=\left|\psi_{n}(t)\right\rangle$. According to the CD driving technology \cite{Berry2009}, the Hamiltonian which reproduces the dynamic evolution
\begin{equation}
i \hbar \partial_{t}\left|\psi_{n}(t)\right\rangle={H}(t)\left|\psi_{n}(t)\right\rangle
\end{equation}
can be constructed by
\begin{equation}
{H}(t)=i \hbar \partial_{t} {V}(t) {V}^{-1}(t).
\end{equation}
After a straightforward derivation, the CD driving Hamiltonian for pseudo-Hermitian Hamiltonian $H$ takes the form
\begin{equation}
\begin{split}
{H}_{tp}(t)=&{H}_{p}(t)+{H}_{1p}(t),\\
{H}_{p}(t)\equiv&\sum_{n}|E_{n}( t )\rangle E_{n}(t)\langle E_{n}^{*}( t )|{U}_n(t),\\
{H}_{1p}(t)\equiv&i \hbar \sum_{n}\left[|\partial_{t} E_{n}(t)\rangle\langle E_{n}^{*}(t)|{U}_n(t)\right.\\
&\left.-\langle E_{n}^{*}(t)|{U}_n(t)| \partial_{t} E_{n}(t)\rangle| E_{n}(t)\rangle\langle E_{n}^{*}(t)| {U}_n(t)\right].
\end{split}
\label{eq:TQDsr}
\end{equation}
If we want to reproduce the adiabatic evolution of $|\psi_n(t)\rangle$, normally we need $H_p(t)$ has full real spectrum, i.e. $|E_n(t)\rangle=|E_n^*(t)\rangle$.  Besides, even the adiabatic condition is satisfied, we also need either real Berry phases by a time independent  $U$ or zero Berry phases. The imaginary part of Berry phase can also cause an unstable evolution.

In many cases, we only need the evolution of $|E_n(t)\rangle$ rather than its dynamic phases and Berry phases. Therefore, we can use the CD part
\begin{equation}
H^p_{CD}(t)=i \hbar \sum_{n}|\partial_{t} E_{n}(t)\rangle\langle E_{n}^{*}(t)|{U}_n(t)
\end{equation}
to realize the CD driving.
\subsection{antipseudo-Hermitian system}
A natural generalization of pseudo-Hermitian system is antipseudo-Hermitian system \cite{[The notition ``antipseudo" here is different with the notition ``anti-pseudo" in ][. We drop the hyphen to distinguish between them.]Mostafazadeh2002bb} whose Hamiltonian satisfies
\begin{equation}
H_{ap}^\dagger=-UH_{ap}U^\dagger.
\label{apH}
\end{equation}This kind of Hamiltonian can be constructed by a pseudo-Hermitian one by
$H_{ap}=iH_p$. $H_{ap}$ shares the same eigenvectors $|E_n\rangle$ with $H_p$ corresponding to the eigenvalues $iE_n$. Therefore, the time dependent eigenvalues of an antipseudo-Hermitian system are pure imaginary or complex ones $E_n(t)$ and $-E^*_n(t)$ which have same imaginary parts and opposite real parts. Its left and right eigenvectors satisfy
\begin{equation}
\langle E_{m}(t)|{U}_n(t)|-E^*_n(t)\rangle=\langle -E^*_m(t)|{U}_n(t)|E_{n}(t)\rangle=\delta_{m n}.
\end{equation}
According to Eq. (\ref{pscon1}), the adiabatic condition
\begin{equation}
\frac{|\left\langle -E^*_n(t)\right|{U}(t)\mid \dot{E}_{m}(t)\rangle|}{\left|\omega_{n m}(t)\right|} e^{-\operatorname{Im}\left[W_{n m}(t)\right]} \ll 1
\end{equation}
for the antipseudo-Hermitian system normally can not be satisfied. The imaginary parts of energy eigenvalues will break the adiabaticity in a vary short time period. Therefore, it is impossible to have an adiabatically evolution of an eigenstate of antipseudo-Hermitian Hamiltonian. Although we can derive the Berry connection and  CD driving Hamiltonian for the antipseudo-Hermitian system
\begin{equation}
\begin{split}
A _{n}(t)&=i\left\langle-E^*_n( t )\left|{U}_{n} (t)\right|\dot{E}_{n}( t )\right\rangle,\\
{H}_{tap}(t)&={H}_{ap}(t)+{H}_{1ap}(t),\\
{H}_{ap}(t)&\equiv\sum_{n}|E_{n}( t )\rangle E_{n}\langle -E^*_n( t )|{U}_n(t),\\
{H}_{1ap}(t)&\equiv i \hbar \sum_{n}\left[|\partial_{t} E_{n}(t)\rangle\langle -E^*_n(t)|{U}_n(t)\right.\\
&\left.-\langle-E^*_n(t)|{U}_n(t)| \partial_{t} E_{n}(t)\rangle| E_{n}(t)\rangle\langle -E^*_n(t)| {U}_n(t)\right]
\end{split}
\label{apeq}
\end{equation}
similar with (\ref{eq:berrysp1}) and (\ref{eq:TQDsr}), the CD driving for $|\psi_n\rangle$ still can not be realized if the dynamic phases and Berry phases are included. We can only accelerate the eigenstate $|E_n\rangle$ by the CD part
\begin{equation}
H^{ap}_{CD}=i \hbar \sum_{n}|\partial_{t} E_{n}(t)\rangle\langle -E_{n}^{*}(t)|{U}_n.
\end{equation}
\subsection{self-normalized energy eigenstate in pseudo- and antipseudo Hermitian system}
Except for the complex energy eigenvalues and Berry phases in non-Hermitian system, the population of an eigenstate of a non-Hermitian Hamiltonian is another  obstacle to hinder its further application. For a right eigenstate $|E_n^r\rangle$ of a non-Hermitian Hamiltonian which satisfies
\begin{equation}
H|E_n^r\rangle=E_n|E_n^r\rangle,
\label{eigennh}
\end{equation}
it can be decomposed by a bare basis $\{|k\rangle\}$ as
\begin{equation}
|E_n^r\rangle=\sum_kC_{kn}|k\rangle.
\end{equation}
The sum of probabilities $\sum_k|C_{nk}|^2=\sum_k\langle k|E_n^r\rangle\langle E_n^r|k\rangle=\langle E_n^r|E_n^r\rangle$ is normally not equal to $1$ since the normalization relation $\langle E_n^l|E_n^r\rangle=1$ in non-Hermitian system requires the left eigenstate $|E_n^l\rangle$. Only if the eigenstate $|E_n^r\rangle$ is self-normalized, i.e. $|E_n^l\rangle=|E_n^r\rangle$, as that for the Hermitian system, the probabilities of bare states are well defined. It is easy to find that the relation between the left and right eigenstate
\begin{equation}
|E_n^l\rangle=U_n|E_n^r\rangle
\end{equation}
for a real (pure imaginary) eigenvalue $E_n$ of pseudo- (antipseudo-) Hermitian Hamiltonian  just provide a way to find a self-normalized energy eigenstate in a non-Hermitian system if $U_n|E_n^r\rangle=|E_n^r\rangle=|E_n\rangle$. By the definition (\ref{eq:dia_1}) and (\ref{apH}), we have
\begin{equation}
\begin{split}
H_p^\dagger|E_n\rangle&=UH_pU^\dagger|E_n\rangle=E_n|E_n\rangle,\\
H_{ap}^\dagger|E_n\rangle&=-UH_{ap}U^\dagger|E_n\rangle=-E_n|E_n\rangle.
\end{split}
\end{equation}
By Eq. (\ref{eigennh}), it becomes
\begin{equation}
\begin{split}
&H^R_{p}|E_n\rangle=E_n|E_n\rangle,~~~~H^I_{p}|E_n\rangle=0,\\
&H^I_{ap}|E_n\rangle=-iE_n|E_n\rangle,~~~~H^R_{ap}|E_n\rangle=0.\\
\end{split}
\label{sncond}
\end{equation}
with $H^R_{p(qp)}=\left[H_{p(ap)}+H^\dagger_{p(ap)}\right]/2$ and $H^I_{p(qp)}=\left[H_{p(ap)}-H^\dagger_{p(ap)}\right]/(2i)$.
This means that an eigenstate is self-normalized in a  pseudo- (antipseudo-) Hermitian system iff it is the common eigenstate for the ``real (imaginary)" part of  pseudo- (antipseudo-) Hermitian with eigenvalue $(-i)E_n$ and the `` imaginary (real)" part of  pseudo- (antipseudo-) Hermitian with eigenvalue $0$. Consider the adiabatic evolution in Eq. (\ref{xde}), the self-normalized energy eigenstates $|\psi_n\rangle$ for antipseudo-Hermitan also require $E_n=0$ (see Apendix \ref{PHS}).
\section{example\label{secv}}
To illustrate the above theoretical results, we now consider a
non-Hermitian stimulated Raman adiabatic passage described by Hamiltonian \cite{PhysRevA.89.063412}
\begin{equation}
{H}_0(t)=\frac12\left(\begin{array}{ccc}
i \gamma_1 & \Omega_{p} & 0 \\
\Omega^*_{p}& i\gamma_2 & \Omega_{s}\\
0 & \Omega_{s}^*& i\gamma_3
\end{array}\right),
\label{model}
\end{equation}
where $\Omega_p$ and $\Omega_s$ are time dependent Rabi frequencies of the pump pulse couples the bare states $|1\rangle$ and $|2\rangle$ and the Stokes pulse couples the bars states $|3\rangle$ and $|2\rangle$. $\gamma_1$, $\gamma_2$ and $\gamma_3$ are the time dependent gain and loss rates.

\begin{figure}[t]
    \centering
    \includegraphics[width=0.48\columnwidth]{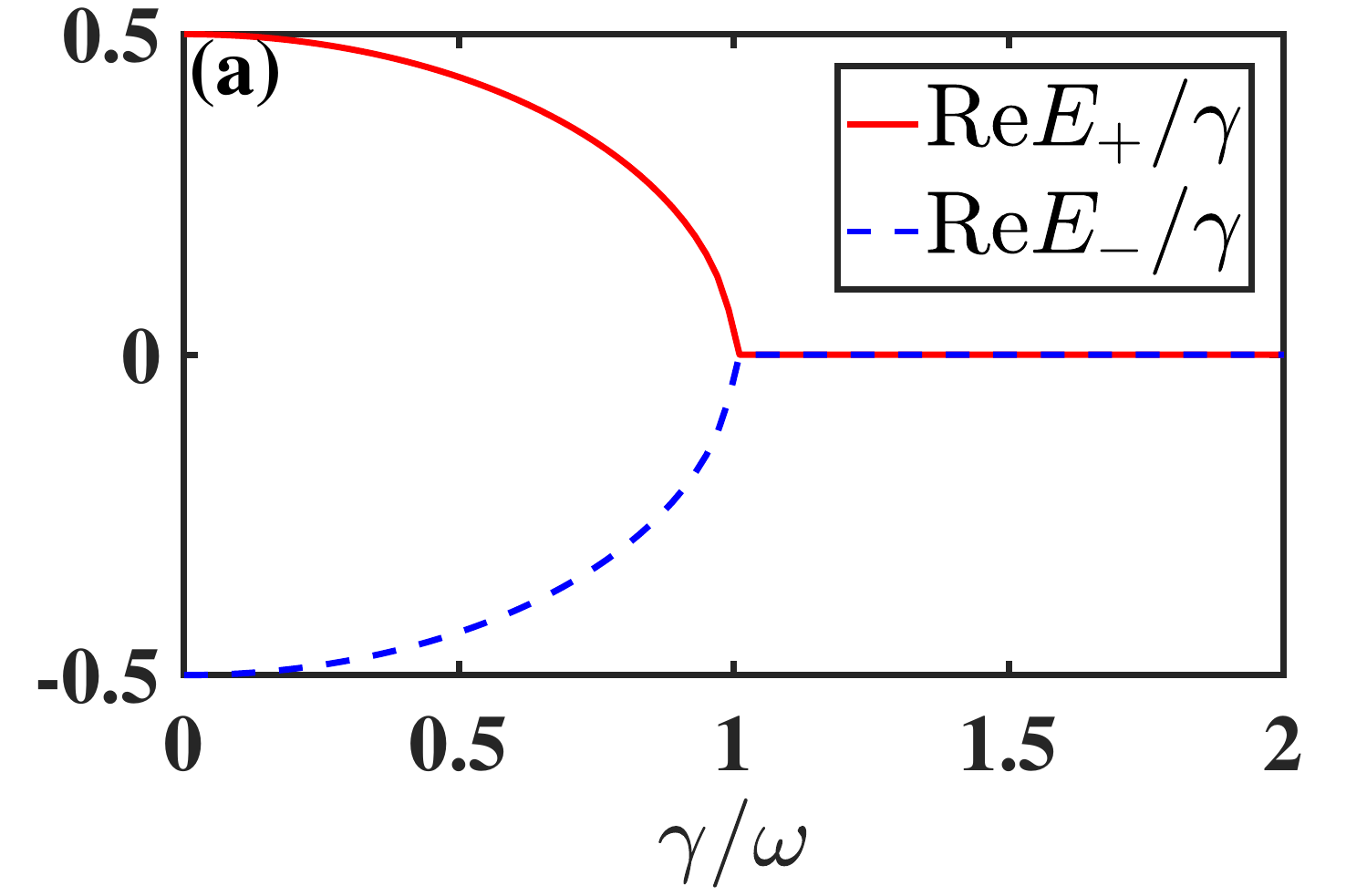}
    \includegraphics[width=0.48\columnwidth]{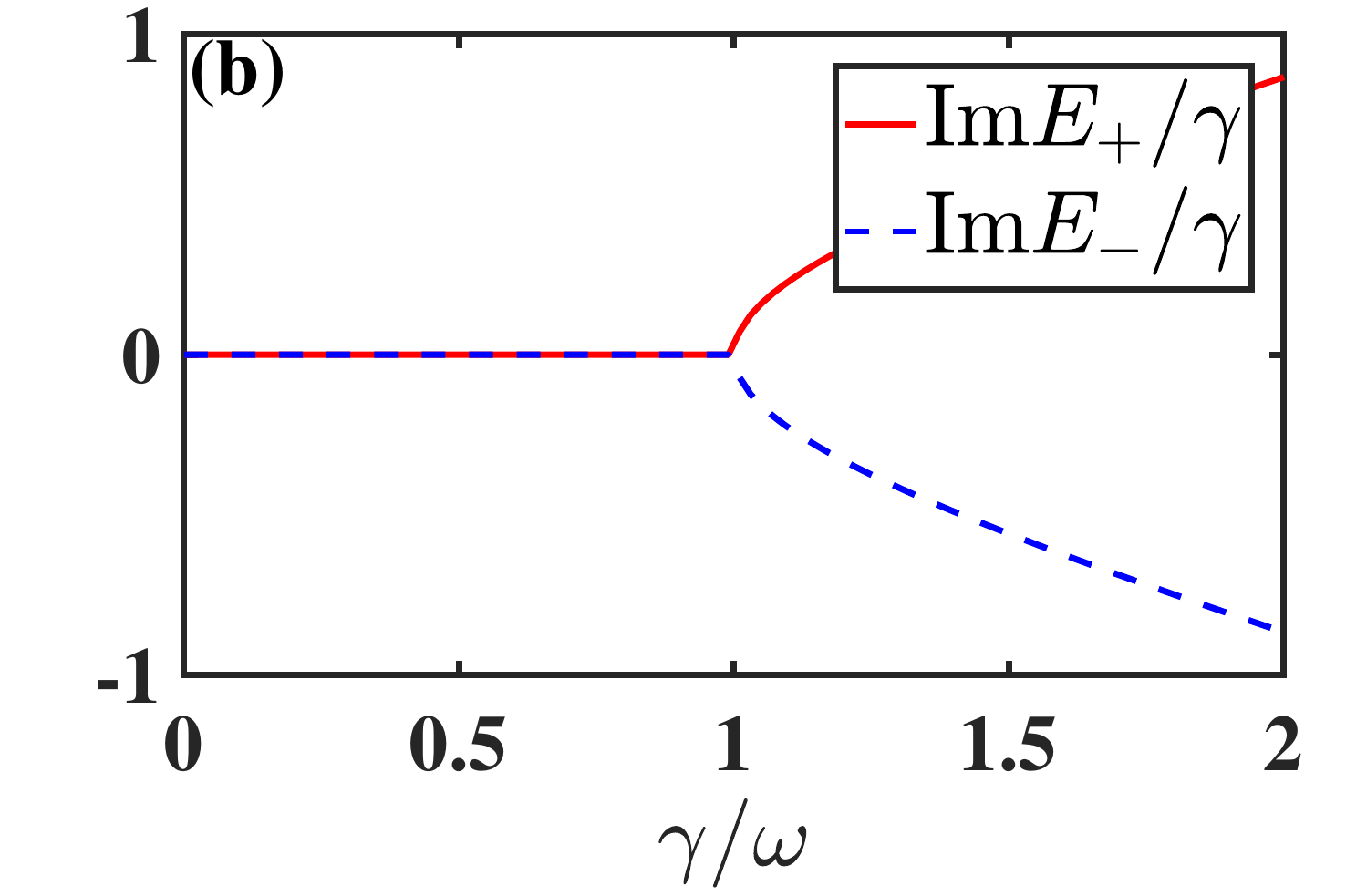}
\caption{Real parts (left panel) and imagine parts (right panel) of energy eigenvalues $E_\pm$ versus $\gamma/\omega$ with $\Omega_0$=1.}
    \label{figph}
\end{figure}

We first consider a pseudo-Hermitian case with $\gamma_1=-\gamma_3=\gamma$, $\gamma_2=0$ and  $\Omega_p=\Omega_s=\frac{1}{\sqrt{2}} \omega e^{ i \varphi} $, the Hamiltonian (\ref{model}) becomes
\begin{equation}
{H}_p(t)=\frac12\left(\begin{array}{ccc}
i \gamma & \frac{1}{\sqrt{2}} \omega e^{ i \varphi} & 0 \\
\frac{1}{\sqrt{2}} \omega e^{- i \varphi}& 0 & \frac{1}{\sqrt{2}} \omega e^{ i \varphi}\\
0 & \frac{1}{\sqrt{2}} \omega e^{ -i \varphi} & -i \gamma
\end{array}\right).
\end{equation}
According to Eq. (\ref{eq:dia_1}), the symmetry matrix $U$ can be chosen as
\begin{equation}
{U}(t)=\left(\begin{array}{ccc}
0 & 0 & e^{2 i \varphi} \\
0 & 1 & 0 \\
e^{-2 i \varphi} & 0 & 0
\end{array}\right).
\end{equation}
The time dependent eigenvalues and eigenstates of this Hamiltonian are
\begin{equation}
E_0=0,E_\pm(t)=\pm\frac{1}{2}\sqrt{\omega^2-\gamma^2}=\pm\Omega_{0}\sqrt{-\cos 2\theta},
\end{equation}
\begin{equation}
|\psi_0(t)\rangle_ =\left(\begin{array}{c}
\frac{-\sqrt{2}\sin \theta}{2\sqrt{-\cos 2 \theta} }  \\
\frac{i\cos \theta e^{-i \varphi}}{\sqrt{-\cos 2 \theta} } \\
\frac{\sqrt{2} \sin \theta e^{-2 i\varphi} }{2\sqrt{-\cos 2 \theta}}
\end{array}\right),
~~|\psi_\pm(t)\rangle_ =\left(\begin{array}{c}
\frac{\sin \theta}{2 \sqrt{-\cos 2 \theta}} \\
\frac{\sqrt{2}(\pm\sqrt{-\cos 2 \theta}-i \cos \theta) e^{-i \varphi}}{2 \sqrt{-\cos 2 \theta}} \\
\frac{(\pm\sqrt{-\cos 2 \theta}-i \cos \theta)^2 e^{-2i \varphi}}{2\sin\theta\sqrt{-\cos2 \theta} } \\
\end{array}\right).
\end{equation}
with $\tan\theta\equiv\omega/\gamma$ and $\Omega_0\equiv\sqrt{\omega^2+\gamma^2}/2$. For  $\omega>\gamma$, the three energy eigenvalues are all real, the three eigenstates satisfy the normalization condition $\left\langle{\psi}_0 (t)| {U}_-(t)|{\psi}_0(t)\right\rangle=1$, $\left\langle{\psi}_\pm(t)| {U}(t)|{\psi}_\pm(t)\right\rangle=1$ with $U_-(t)=-U(t)$. For $\omega<\gamma$, the two of energy eigenvalues become a pair of complex conjugates, the normalization condition becomes  $\left\langle{\psi}_0 (t)| {U}(t)|{\psi}_0(t)\right\rangle=1$, $\left\langle{\psi}_+(t) | {U} _{-} (t)|{\psi}_-(t)\right\rangle=1$. This is worth to note that the energy spectrum of this pseudo-Hermitian system possess a structure of EPs as shown in Fig. \ref{figph}, since it is also a typical PT-symmetric systems \cite{Bender_2007} with gain and loss.

\begin{figure}[t]
\centering
 \includegraphics[width=0.48\columnwidth]{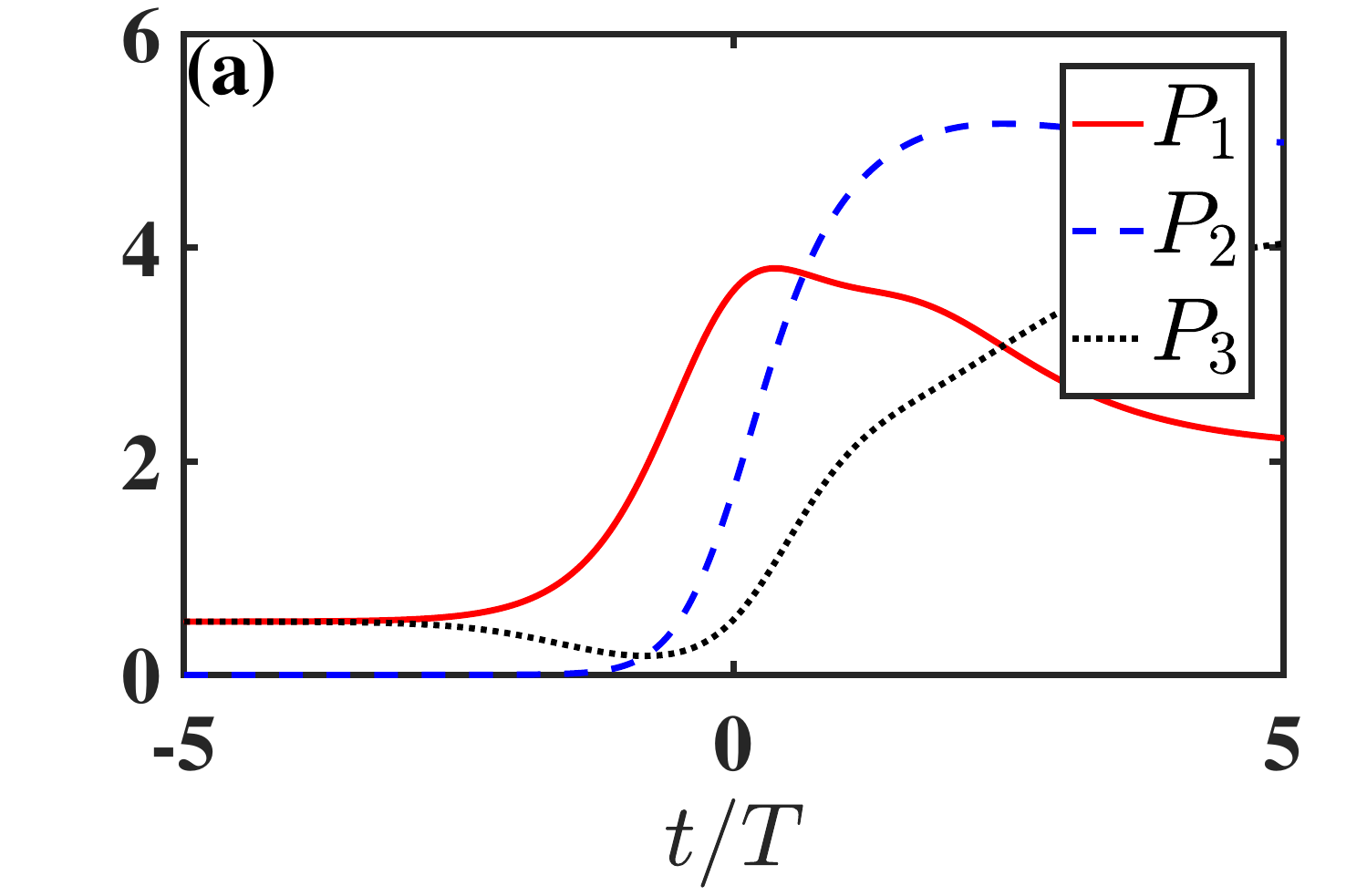}
    \includegraphics[width=0.48\columnwidth]{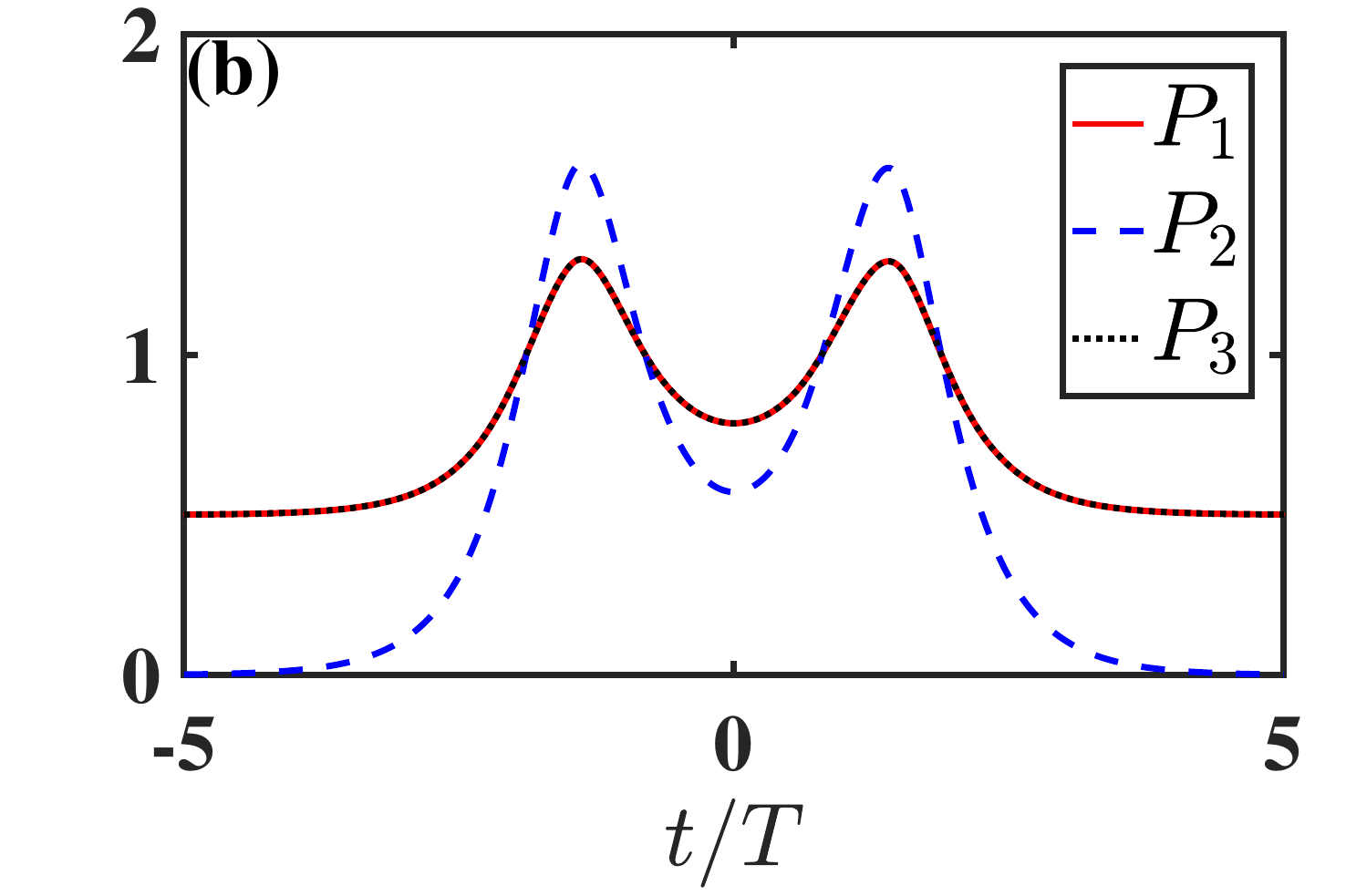}
     \includegraphics[width=0.48\columnwidth]{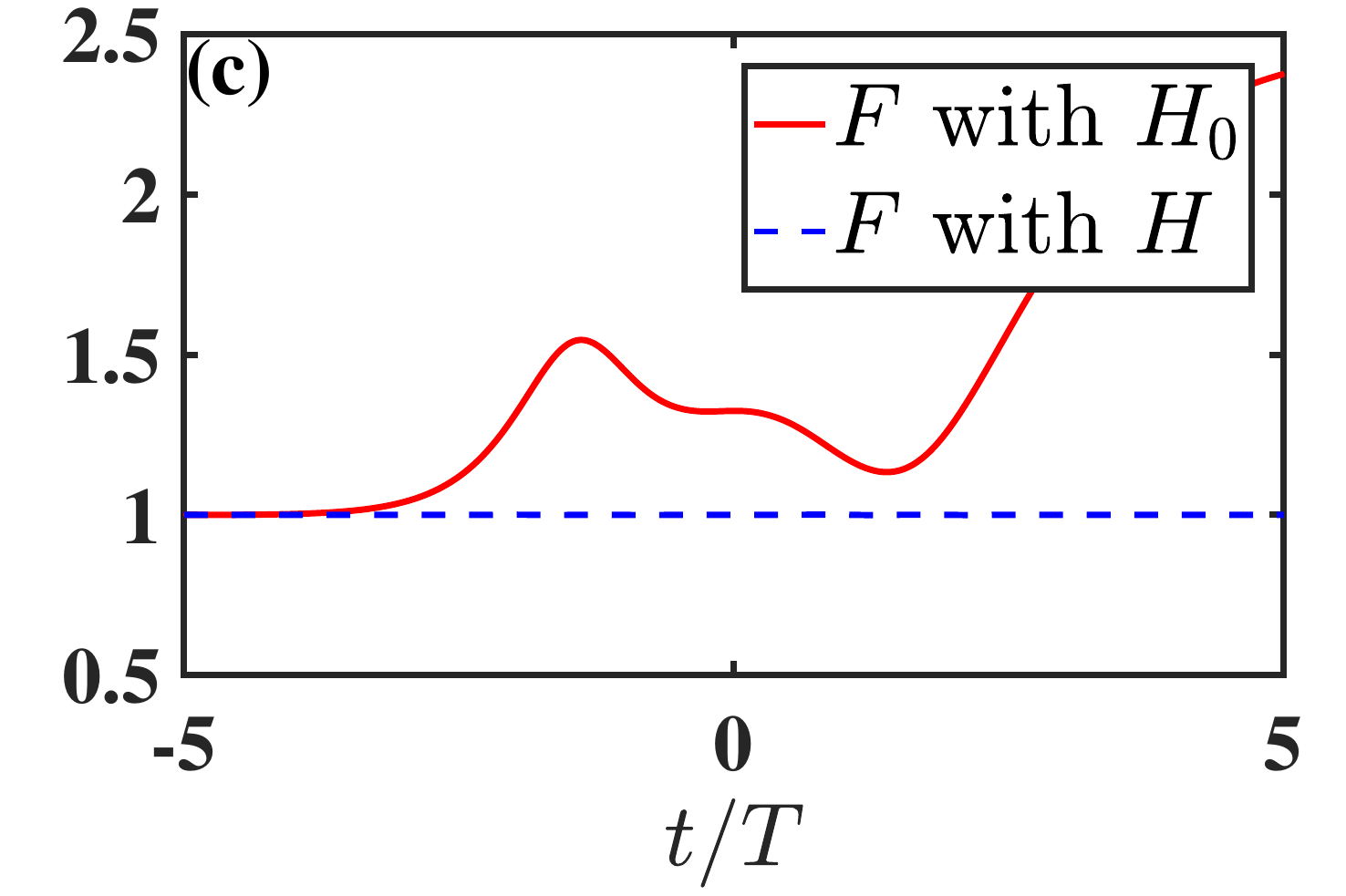}
    \includegraphics[width=0.48\columnwidth]{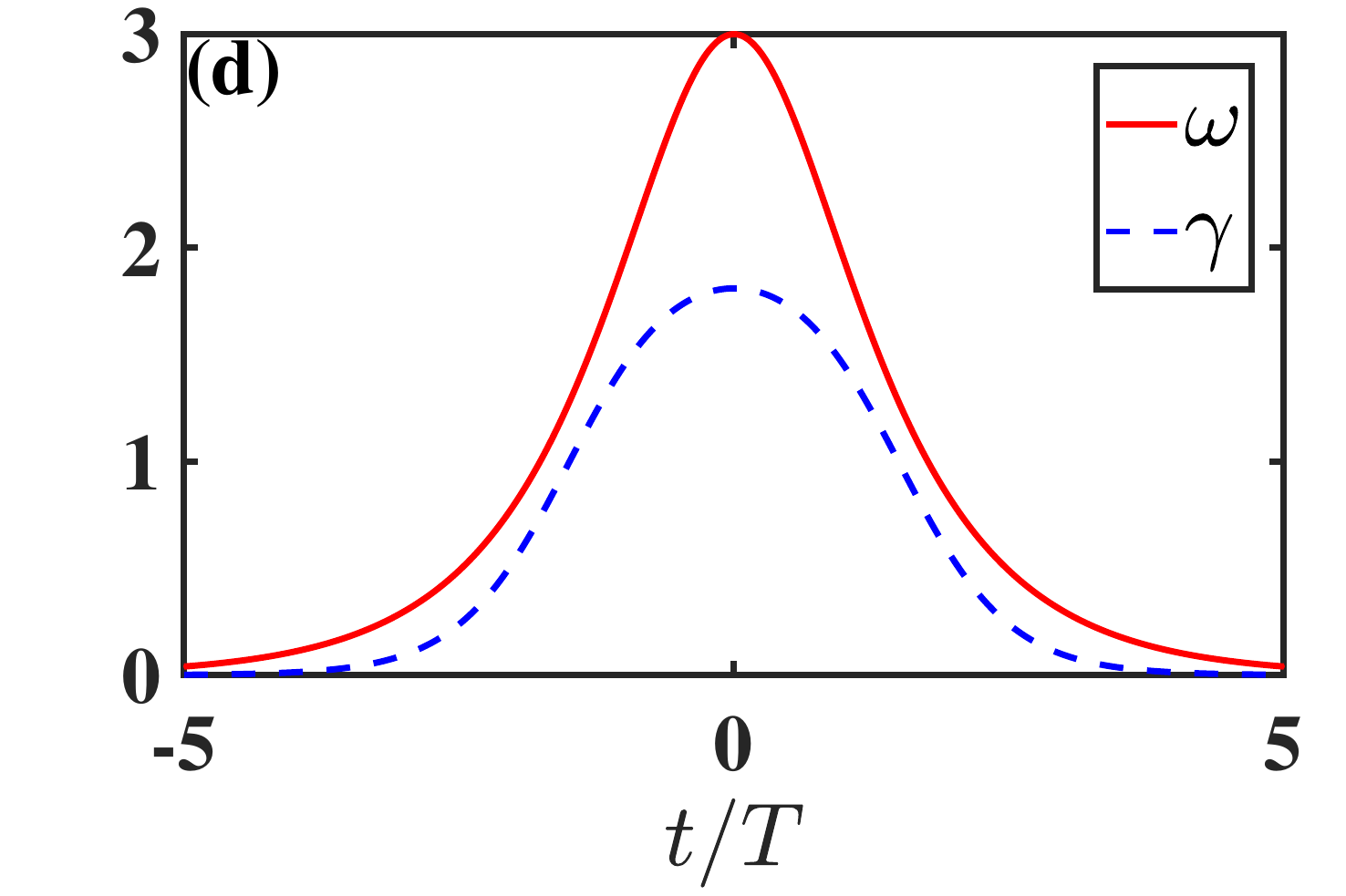}
\caption{(Color online) Time evolution of the Populations on $|1\rangle$ (solid red line), $|2\rangle$ (dashed blue line), and $|3\rangle$ (black dotted line) driven by (a) the original Hamiltonian $H_{p}(t)$ and (b) the CD driving Hamiltonian $H_{tp}(t)$; (c) Time evolution of the fidelities on $|\psi_0(0)\rangle$; (d) Shapes of $\omega$ and $\gamma$. The parameters are chosen as $T=1/\Omega_0$.}
\label{fig:realph}
\end{figure}

Since the eigenstates depend on two parameters $\theta$ and $\phi$, the Berry connection can be written as
\begin{equation}
A_n=A_{\theta_n}\dot{\theta}+A_{\varphi_n}\dot{\varphi}
\end{equation}
with $A_{\theta_n}\equiv i\langle \psi_n^*(t)|U_n(t)\partial_{\theta}|\psi_n(t)\rangle$, $A_{\varphi_n}\equiv i\langle \psi_n^*(t)|U_n(t)\partial_{\varphi}|\psi_n(t)\rangle$.
According to Eq.(\ref{eq:berrysp1}), we have
\begin{equation}
\begin{split}
A_{\theta_ 0}&=0,\quad
A_{\theta_ +}=-\frac{1}{\sin \theta \sqrt{-\cos 2 \theta}},\quad
A_{\theta_ -}=\frac{1}{\sin \theta \sqrt{-\cos 2 \theta}},\\
A_{\varphi_ 0}&=1,\quad
A_{\varphi_ +}=1-\frac{i\cos \theta}{\sqrt{-\cos 2 \theta}},\quad
A_{\varphi_ -}=1+\frac{i\cos \theta}{\sqrt{-\cos 2 \theta}}.
\end{split}
\end{equation}
For the real spectrum ($\omega>\gamma$), the Berry connections $A_{\theta_n}$ are all real while  $A_{\varphi_\pm}$ are complex. Notice that the time integration of  $A_{\theta_n}\dot{\theta_n}$ can be transformed by an integration of $\theta$ which will vanish in a cyclic evolution.
Under the method of the CD driving, the Hamiltonian takes the form
\begin{equation}
H_{tp}(t)=H_p(t)+H_{1p}(t)
\label{ex1H}
\end{equation}
with the additional Hamiltonian
\begin{equation}
H_{1p}(t)=\left(\begin{array}{ccc}
\frac{\sin^{2} \theta}{\cos 2 \theta} \dot{\varphi} & \frac{-e^{i \varphi}(2\dot{\theta}+i\dot{\varphi}\sin2\theta)}{2\sqrt{2} \cos 2 \theta} & 0 \\
\frac{e^{-i \varphi}(2\dot{\theta}-i\dot{\varphi}\sin2\theta)}{2\sqrt{2} \cos 2 \theta} & 0 & \frac{-e^{i \varphi}(2\dot{\theta}+i\dot{\varphi}\sin2\theta)}{2\sqrt{2} \cos 2 \theta} \\
0 & \frac{e^{-i \varphi}(2\dot{\theta}-i\dot{\varphi}\sin2\theta)}{2\sqrt{2} \cos 2 \theta} & \frac{-\sin^{2} \theta}{\cos 2 \theta} \dot{\varphi}
\end{array}\right).
\end{equation}
Next we show the evolution of the populations $P_k(t)=|C_{0k}(t)|^2$ of the bare states $|k\rangle$ on the eigenstate $|\psi_0(t)\rangle=\sum_kC_{0k}(t)|k\rangle$ to test the effect of the CD driving. For the case of full real spectrum ($\omega>\gamma$), we take the parameters
\begin{equation}
\begin{split}
\omega&=3\Omega_{0}\operatorname{sech}(t / T),\\
\gamma&=[\operatorname{tanh}(t/T+3 / 2)-\operatorname{tanh}(t/T-3/ 2)]/T
\end{split}
\label{ex1set}
\end{equation}
and $\varphi=0$, the shape of parameters are shown in Fig. \ref{fig:realph}d. As shown in  Fig. \ref{fig:realph}a-c, the CD driving Hamiltonian $H_{tp}(t)$ in Eq. (\ref{ex1H})  can perfectly reproduce the adiabatic evolution of $|\psi_0(t)\rangle$ with a high fidelity ( the fidelity of $|\psi(t)\rangle$ on $|\psi_0(t)\rangle$ are defined by $F=|\langle\psi(t)|U(t)|\psi_0(t)\rangle|$) rather than the original Hamiltonian $H_p(t)$.

\begin{figure}[t]
\centering
 \includegraphics[width=0.48\columnwidth]{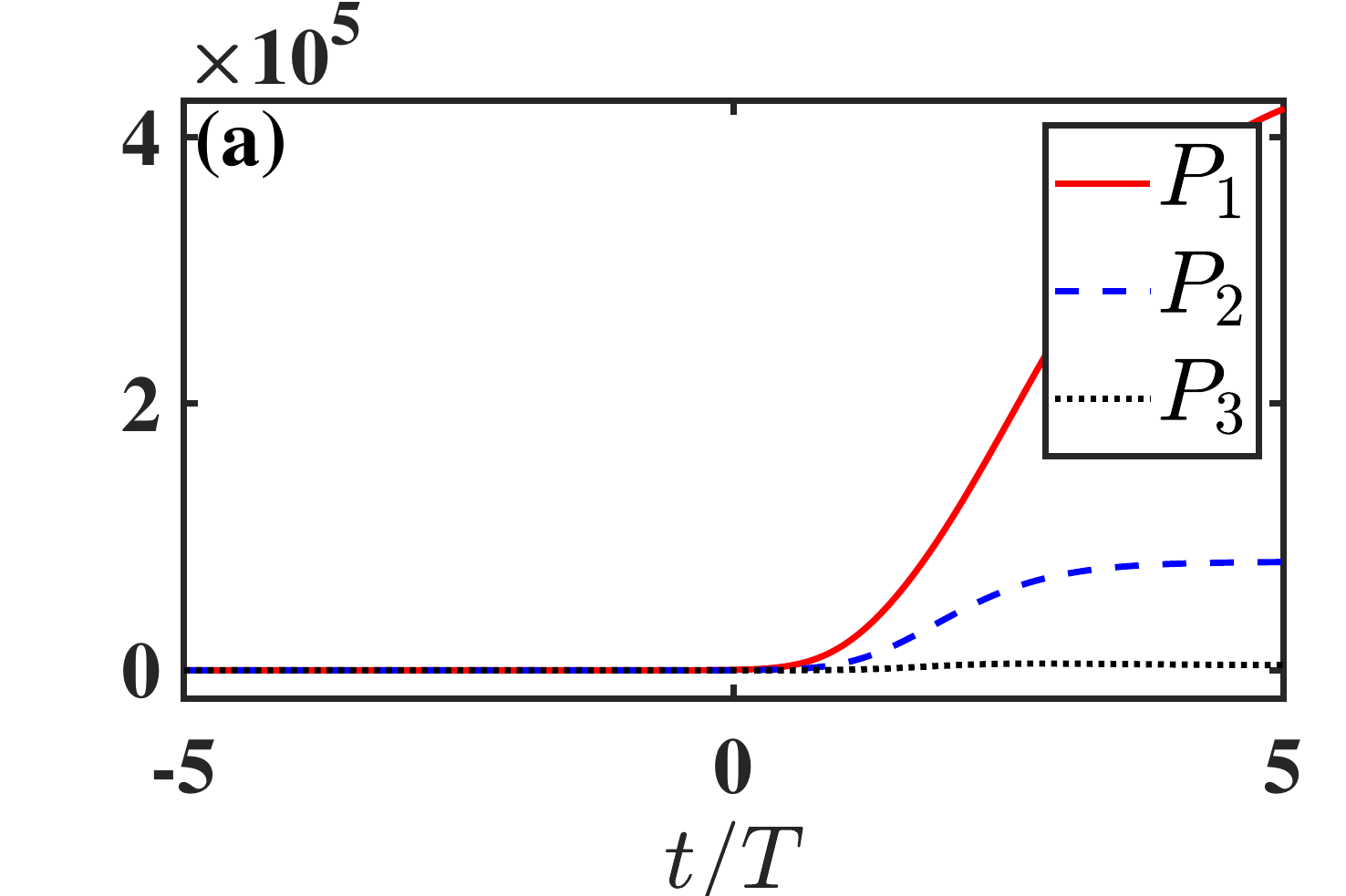}
    \includegraphics[width=0.48\columnwidth]{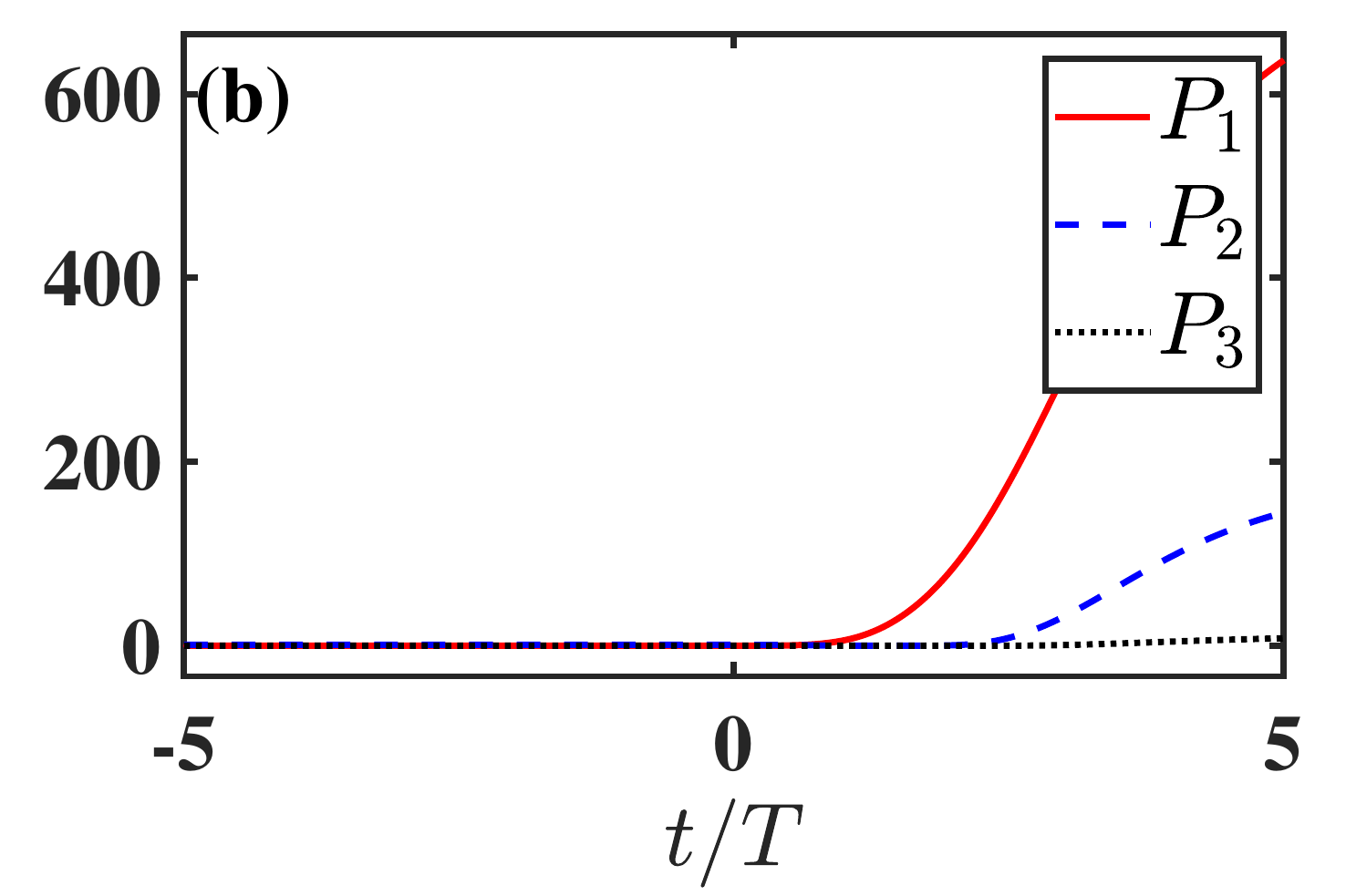}
     \includegraphics[width=0.48\columnwidth]{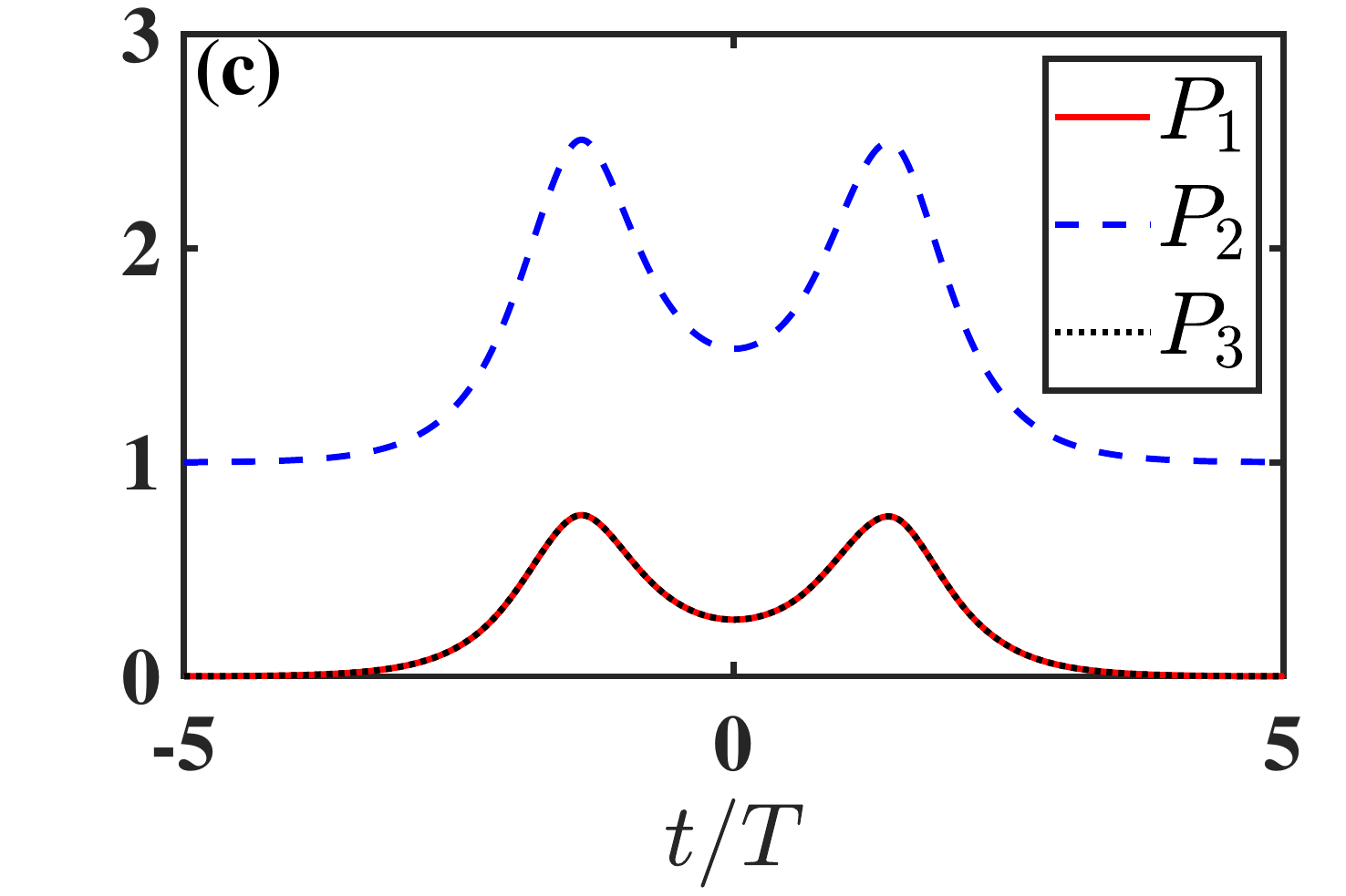}
    \includegraphics[width=0.48\columnwidth]{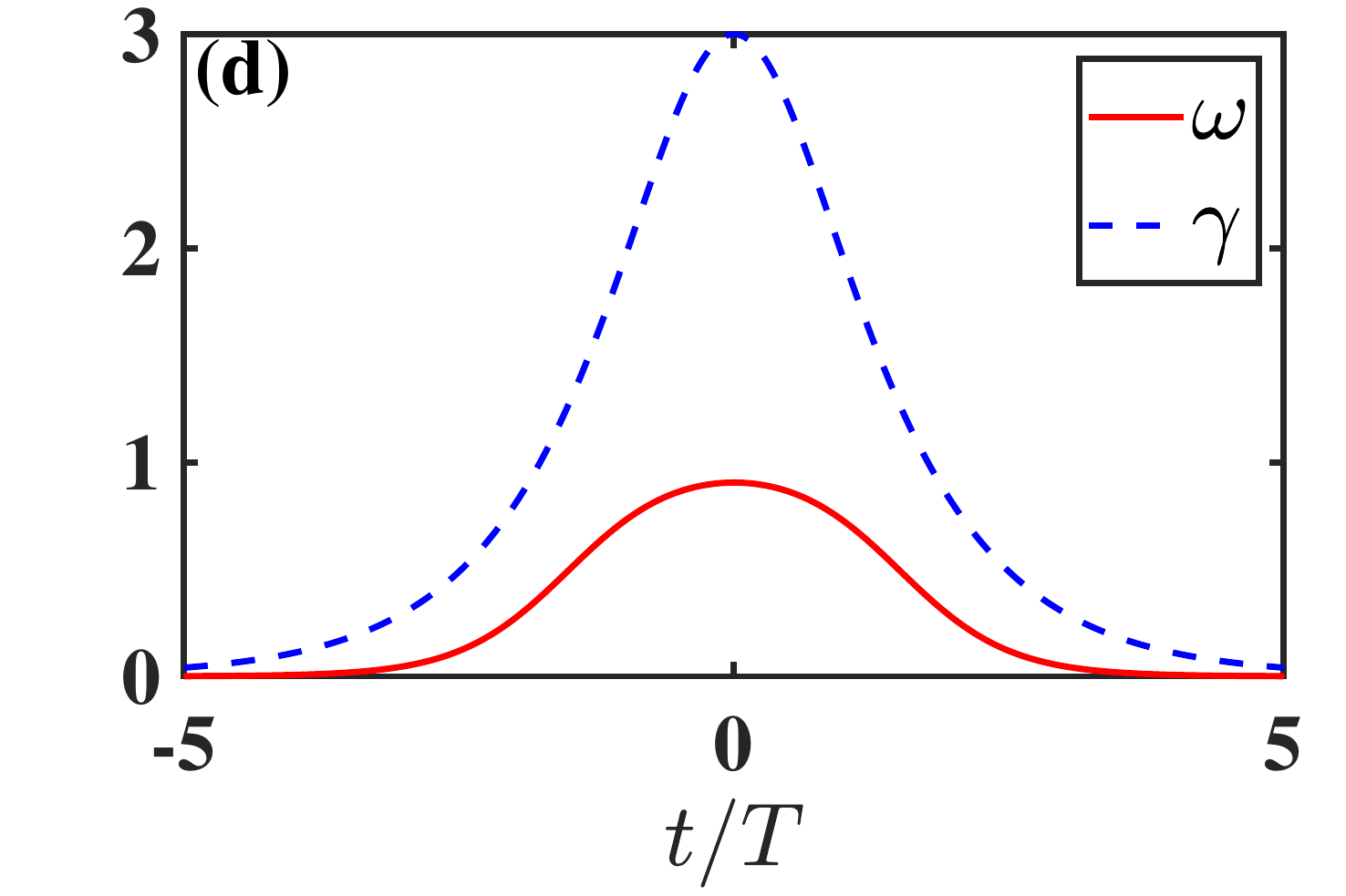}
\caption{(Color online) Time evolution of the Populations on $|1\rangle$ (solid red line), $|2\rangle$ (dashed blue line), and $|3\rangle$ (black dotted line) driven by (a) the original Hamiltonian $H_{p}(t)$, (b) the CD driving Hamiltonian $H_{tp}(t)$ and (c) the additional Hamiltonian$H_{CD}^p(t)$; (d) Shapes of $\omega$ and $\gamma$. The initial state is chosen as $|\psi_0(0)\rangle$ and the parameters are chosen as $T=2/\Omega_0$.}
\label{fig:imagph}
\end{figure}

For the case of complex spectrum ($\omega<\gamma$), we switch the setups of the two parameters in Eq. (\ref{ex1set}), the shape of parameters are shown in Fig. \ref{fig:imagph}d. As shown in Fig. \ref{fig:imagph}, neither the CD driving Hamiltonian  $H_{tp}(t)$ nor the original Hamiltonian $H_p(t)$ can reproduce the adiabatic evolution of $|\psi\rangle$. This is caused by the instabilities brought by the complex eigenvalues and Berry phases. In this case, the adiabatic evolution for  $|\psi_0(t)\rangle$ is not exist. To derive the exact evolution of $|\psi_0(t)\rangle$, it needs to be driven by the CD part of $H_{tp}(t)$
\begin{equation}
H_{CD}^p(t)=\left(\begin{array}{ccc}
\frac{i\cos \theta}{\sin \theta\cos 2 \theta} \dot{\varphi} & 0 & 0 \\
\frac{\sqrt2e^{-i \varphi}}{\cos 2 \theta}\dot{\theta} & \dot{\varphi} & 0 \\
0 & \frac{\sqrt2e^{-i \varphi}}{\cos 2 \theta}\dot{\theta} & 2\dot{\varphi}-\frac{i\cos \theta}{\sin \theta\cos 2 \theta} \dot{\theta}
\end{array}\right)
\end{equation}
as shown in Fig. \ref{fig:imagph}c.

Although we can find ways to realize the CD driving for both the cases of real spectrum and complex spectrum, the populations of bare states are not well defined. As shown in Fig. \ref{fig:realph} and Fig. \ref{fig:imagph}, the populations can even larger than $1$. This is caused by the biorthonormal nature of the left and right eigenstates in the non-Hermitian system. Next, we consider an antipseudo-Hermitian case which has self-normalized energy eigenstates to realize population transfer of the bare states.

\begin{figure}[t]
    \centering
    \includegraphics[width=0.48\columnwidth]{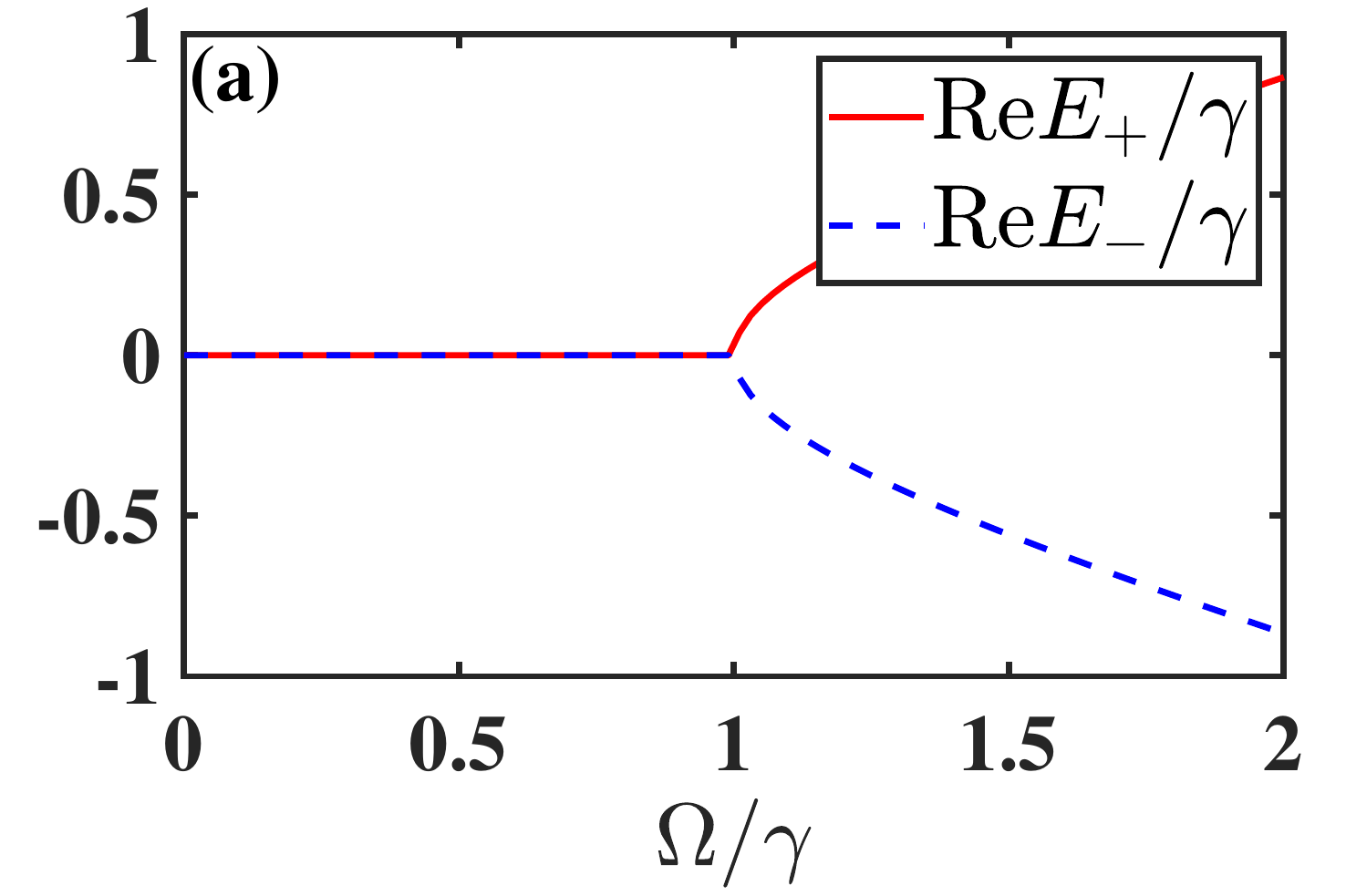}
    \includegraphics[width=0.48\columnwidth]{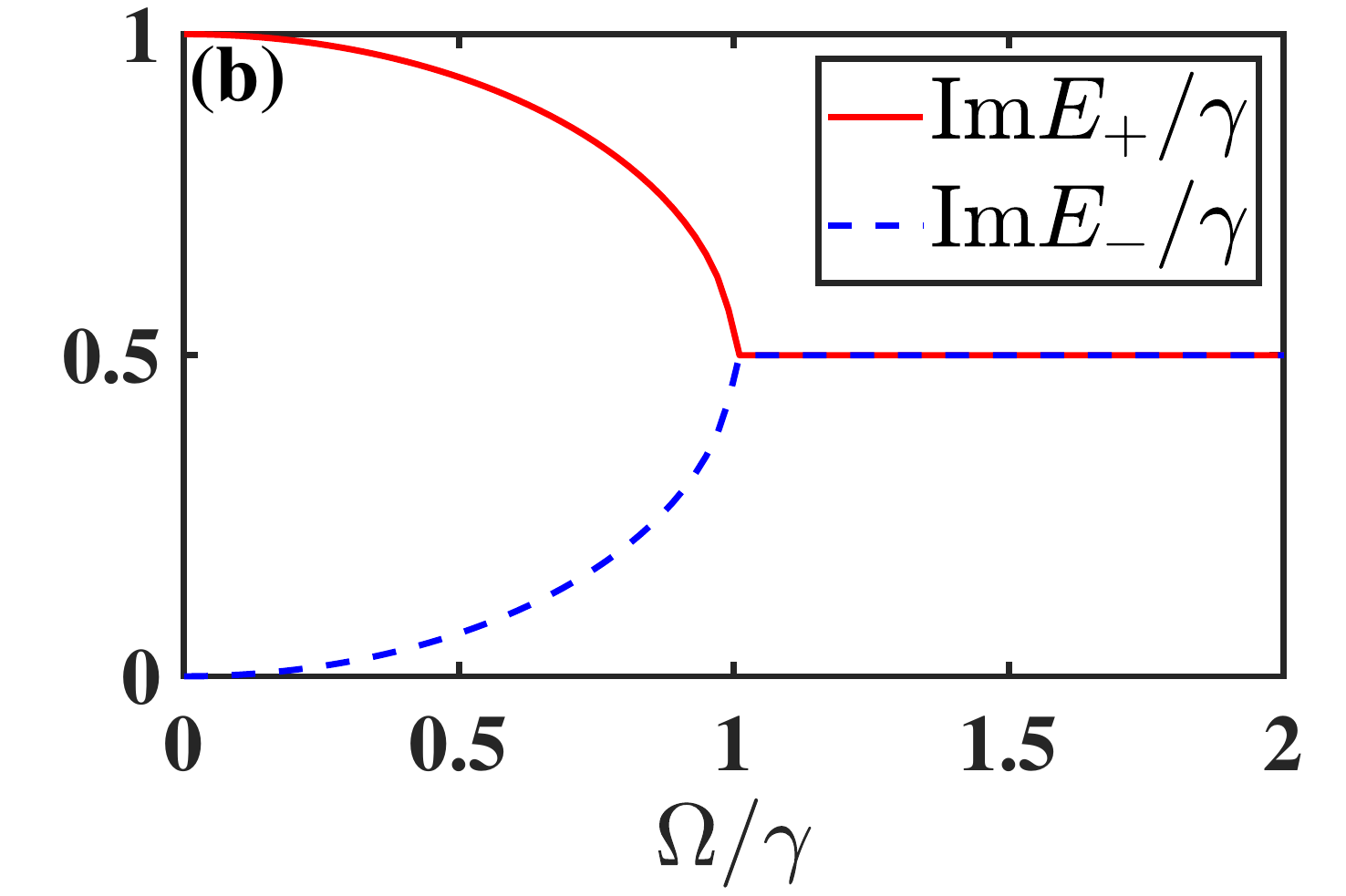}
\caption{(a) Real parts and (b) imagine parts of energy eigenvalues $E_\pm$ versus $\omega/\gamma$.}
    \label{fig:antiE}
\end{figure}

By setting $\gamma_1=\gamma_3=0$ and $\gamma_2=2\gamma$ in Eq. (\ref{model}), we can derive an antipseudo-Hermitian Hamiltonian by
\begin{equation}
{H}_{ap}(t)=\frac12\left(\begin{array}{ccc}
0 & \Omega_{p} & 0 \\
\Omega_{p} & 2i\gamma & \Omega_{s} \\
0 & \Omega_{s} & 0
\end{array}\right).
\end{equation}
According to Eq. (\ref{eq:dia_1}), this Hamiltonian is an antipseudo-Hermitian one. The symmetry matrix $U$ can be chosen as
\begin{equation}
{U}=\left(\begin{array}{ccc}
1 & 0 & 0 \\
0 & -1 & 0 \\
0 & 0 & 1
\end{array}\right).
\end{equation}
The eigenvalues and eigenstates of this Hamiltonian are
\begin{equation}
E_0=0,E_\pm(t)=\frac i2\left(\gamma\pm\sqrt{\gamma^2-\Omega^2}\right),
\end{equation}
\begin{equation}
|\psi_0(t)\rangle_ =\left(\begin{array}{c}
\cos\theta \\
0\\
-\sin\theta
\end{array}\right),
~~|\psi_\pm(t)\rangle_ =\left(\begin{array}{c}
\frac{\sin \theta\sqrt{\cos\phi\mp\sqrt{\cos2\phi}}}{(4\cos2\phi)^{1/4}} \\
\frac{i\sqrt{\cos\phi\pm\sqrt{\cos2\phi}}}{(4\cos2\phi)^{1/4}} \\
\frac{\cos\theta\sqrt{\cos\phi\mp\sqrt{\cos2\phi}}}{(4\cos2\phi)^{1/4}}  \\
\end{array}\right)
\end{equation}
with $\tan\theta\equiv\Omega_p/\Omega_s$, $\Omega\equiv\sqrt{\Omega_p^2+\Omega_s^2}$ and $\tan\phi=\Omega/\gamma$. The two energy eigenvalues $E_\pm$ are either both imaginary (for  $\gamma>\Omega$,) or have opposite real parts. The three eigenstates satisfy the normalization condition
\begin{equation}
\begin{aligned}
&\left\langle{\psi}_0 (t)| {U}|{\psi}_0(t)\right\rangle=\left\langle{\psi}_0 (t)|{\psi}_0(t)\right\rangle=1,\\
&\left\{\begin{array}{ll}
\left\langle{\psi}_\pm(t)| {U}_\mp|{\psi}_\pm(t)\right\rangle=1, & \gamma>\Omega\\
\left\langle{\psi}_+ (t)| {U} |{\psi}_-(t)\right\rangle=1, & \gamma<\Omega
\end{array}\right.
\end{aligned}
\end{equation}
with $U_\pm=\pm U$. Like the pseudo-Hermitian case, the energy spectrum of the antipseudo-Hermitian Hamiltonian $H_{ap}$ also possesses a structure of EPs as shown in Fig. (\ref{fig:antiE}). It is easy to find that the eigenstate $|\psi_0(t)\rangle$
is  a common eigenstate for the ``real" and ``imaginary" parts
\begin{equation}
{H}^R_{ap}(t)=\frac12\left(\begin{array}{ccc}
0 & \Omega_{p} & 0 \\
\Omega_{p} & 0 & \Omega_{s} \\
0 & \Omega_{s} & 0
\end{array}\right),~~~~{H}^I_{ap}(t)=\frac12\left(\begin{array}{ccc}
0 & 0 & 0 \\
0 & 2\gamma & 0 \\
0 & 0 & 0
\end{array}\right)
\end{equation}
of ${H}_{ap}(t)$ with eigenvalues $0$. Therefore, $|\psi_0(t)\rangle$ is an example of the self-normalized energy eigenstates in non-Hermitian system.

By Eq. (\ref{apeq}), the Berry connections can be derived by
\begin{equation}
A_n=0.
\end{equation}
This means that the eigenstates will not accumulate geometric phase via the dynamic evolution.
Under the method of the CD driving in Eq. (\ref{apeq}), the Hamiltonian takes the form
\begin{equation}
H_{CD}^{ap}=H_{ap}+H_{1ap},
\end{equation}
with the additional Hamiltonian
\begin{equation}
H_{1ap}=H_{CD}^{ap}=\left(\begin{array}{ccc}
0 & \frac{\sin\theta}{2 \cos 2 \phi} \dot{\phi} & i\dot{\theta} \\
-\frac{\sin\theta}{2 \cos 2 \phi} \dot{\phi} & 0 & -\frac{\cos\theta}{2 \cos 2 \phi} \dot{\phi}\\
-i\dot{\theta} & \frac{\cos\theta}{2 \cos 2 \phi} \dot{\phi} & 0
\end{array}\right).
\end{equation}

Next, we show the CD driving in antipesudo-Hermitian system by setting
the parameters
\begin{equation}
\begin{split}
\Omega_1&=5\Omega_{0}\operatorname{sech}(t / T-3/2)\\
\Omega_2&=5\Omega_{0}\operatorname{sech}(t / T+3/2)\\
\gamma=&[\operatorname{tanh}(t/T+3 / 2)-\operatorname{tanh}(t/T-3 / 2)]/T
\end{split}
\end{equation}
\begin{figure}[t]
\centering
 \includegraphics[width=0.48\columnwidth]{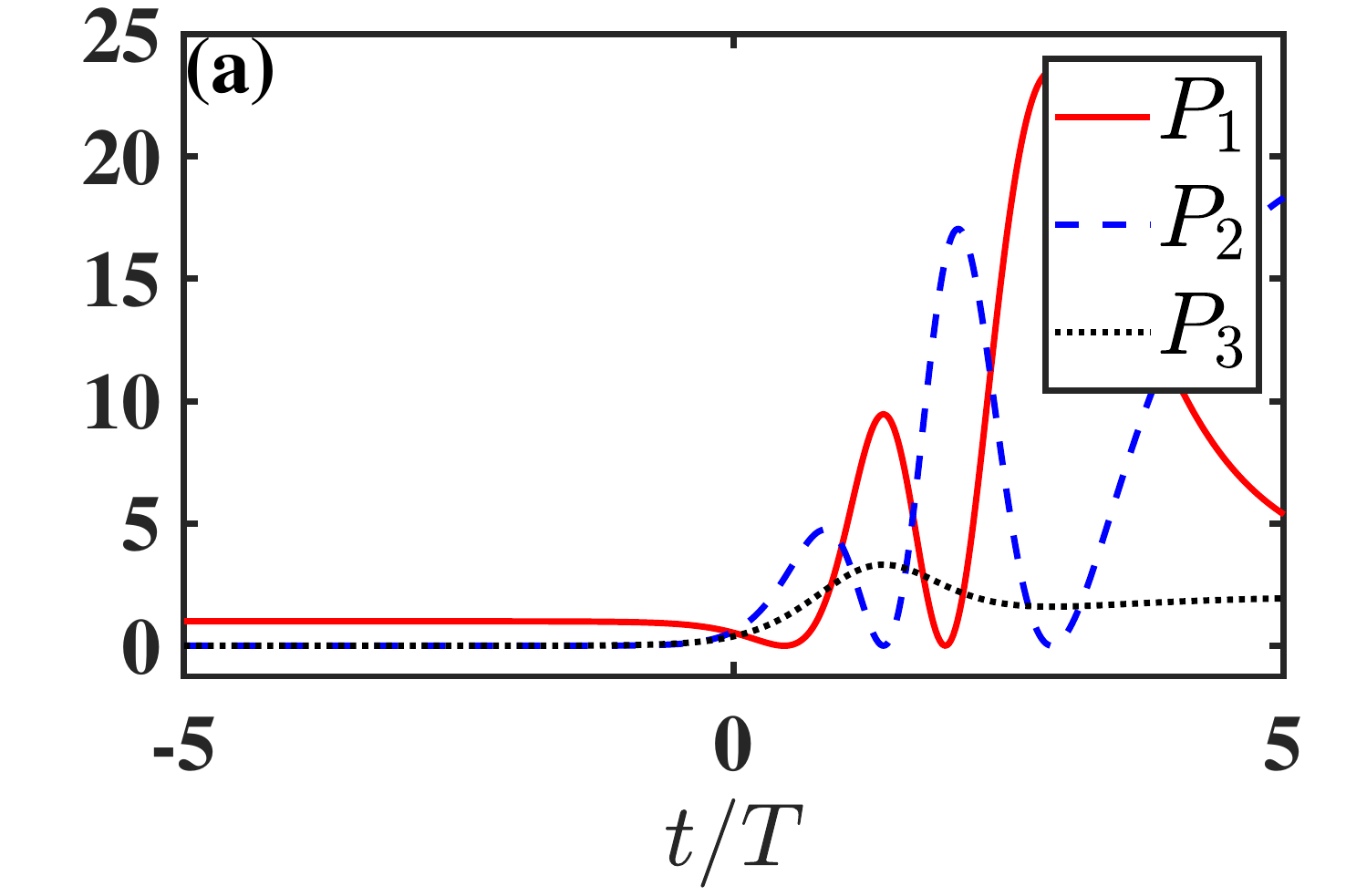}
    \includegraphics[width=0.48\columnwidth]{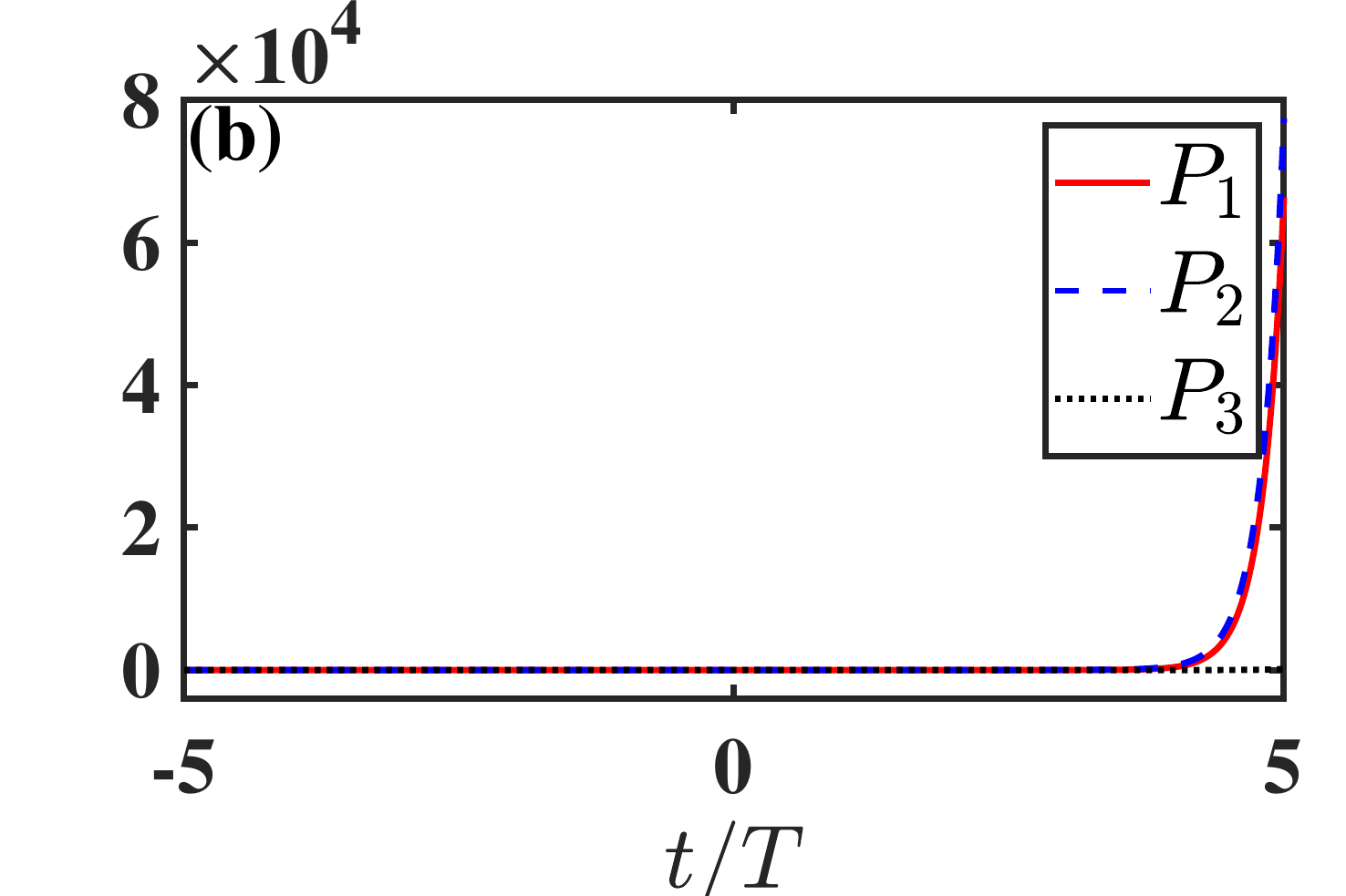}
     \includegraphics[width=0.48\columnwidth]{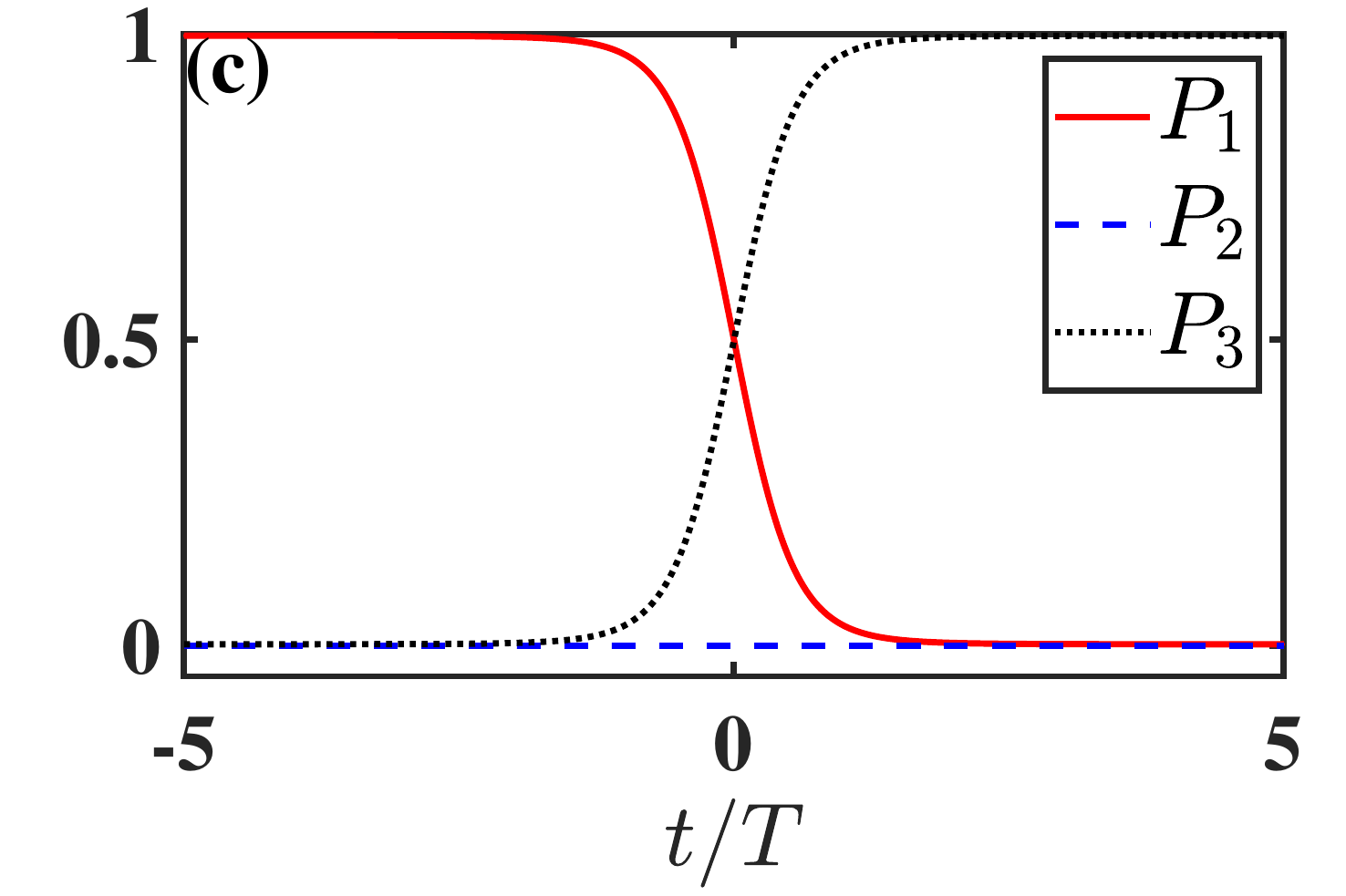}
    \includegraphics[width=0.48\columnwidth]{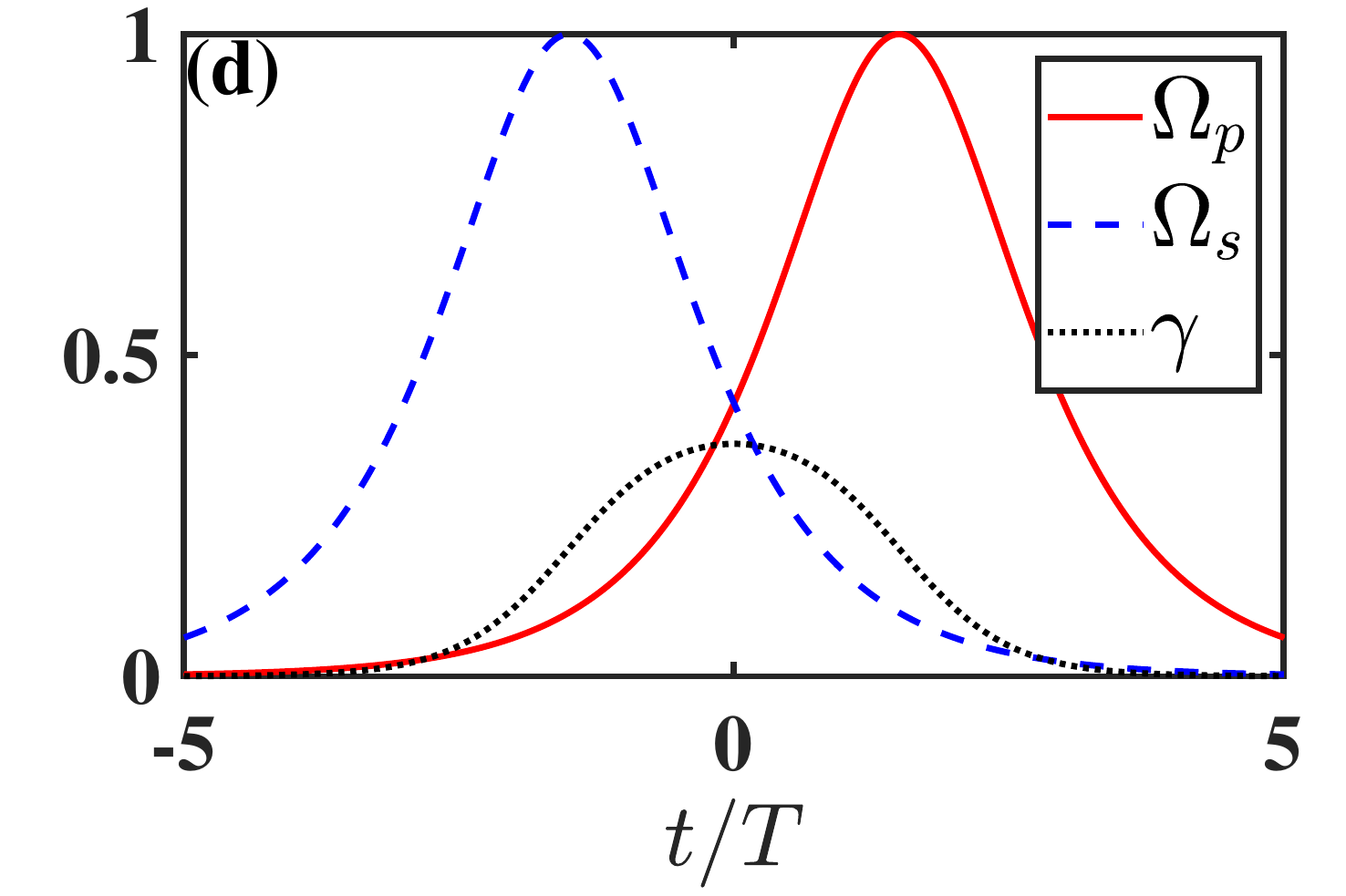}
\caption{(Color online) Time evolution of the Populations on $|1\rangle$ (solid red line), $|2\rangle$ (dashed blue line), and $|3\rangle$ (black dotted line) driven by (a) the original Hamiltonian $H_{ap}$, (b) the CD driving Hamiltonian $H_{CD}^{ap}$ and (c) the Hamiltonian $H_{1ap}$; (d) Shapes of $\omega$ and $\gamma$. The initial state is chosen as $|\psi_0\rangle$ and the parameters are chosen as $T=5/\Omega_0$.}
\label{fig:antimag}
\end{figure}
As shown in Fig. \ref{fig:antimag}, both of the  CD driving Hamiltonian  $H_{tap}(t)$ and the original Hamiltonian $H_{ap}(t)$ fail to reproduce the adiabatic evolution of $|\psi\rangle$. Like the case of pseudo-Hermitian with complex energy eigenvalues, there is no adiabatic evolution exist in antipesudo-Hermitian system since its eigenvalues are always imaginary or complex. However, we can derive the exact evolution of $|\psi_0(t)\rangle$ by the Hamiltonian $H_{1ap}$ and realize perfect population transfer from the bare state $|1\rangle$ to $|3\rangle$ as shown in Fig. \ref{fig:antimag}c. Interestingly, this CD driving process is driven by a non-Hermitian Hamiltonian $H_{1ap}$. This means that we can realize the STA of a self-normalized state by a non-Hermitian Hamiltonian.
\section{ Conclusion\label{secvi}}
In summary, we have studied the CD driving scheme for STA in pseudo- and antipseudo- Hermitian system. By discussing the adiabatic condition for non-Hermitian system, we show that only in the non-Hermitian system which possesses real energy spectrum, its energy eigenstates can adiabatically evolve. Therefore, the adiabatic evolution with dynamic phases and Berry phases can only reproduced by the CD driving in the non-Hermitian system with real spectrum, otherwise the parts of dynamic phases and Berry phases in the CD Hamiltonian should by dropped to realize the exact evolution of an energy eigenstate in a non-Hermitian system. In this sense, we derive the adiabatic conditions and  CD driving Hamiltonian for the antipseudo-Hermitian Hamiltonian which possesses either real or complex energy spectrum and  the pseudo-Hermitian Hamiltonian which possesses either imaginary or complex energy spectrum. We also find that these two kinds of non-Hermitian Hamiltonian naturally provide a way to find self-normalized energy eigenstates in non-Hermitian systems. The population of bare states on this energy eigenstate are well-defined which are normally absent in the non-Hermitian system.
This means that we can realize the CD driving of a self-normalized state by a non-Hermitian Hamiltonian. By using an example, we illustrate our results and realize the perfect population transfer with loss or gain. Our theory can be expected to find applications in realizing STA in the non-Hermitian systems.
\section*{Acknowledgments}

We thank L. B. Fu, F. Q. Dou, X. Guo and C. G. Liu for help discussions. This work is supported by National Natural Science Foundation
of China (NSFC) (Grants Nos. 11875103, 12147206 and 11775048).

\appendix
\section{the projective Hilbert space for the non-unitary time evolution under
non-Hermitian Hamiltonians\label{PHS}}
consider the Schr{\"o}dinger equation
\begin{equation}
i\hbar\dot\psi(t)=H(t)\psi(t)
\end{equation}
with a non-Hermitian Hamiltonian $H$. The state $|\psi(t)\rangle$ normally can not be normalized since  $\langle\psi(t)|\psi(t)\rangle$ is time dependent. However, we can define $|\psi(t)\rangle=e^{\alpha(t)+i\beta(t)}|\tilde\psi(t)\rangle$ to grantee $\langle\tilde\psi(t)|\tilde\psi(t)\rangle=1$. The real coefficients $\alpha(t)$ and $\beta(t)$ satisfy \cite{Fu2020}
\begin{equation}
\begin{aligned}
\dot\alpha(t)=\bar{H}_I(t)/\hbar,~~~~\dot\beta(t)=-\bar{H}_R(t)/\hbar+i\langle\tilde\psi(t)|\dot{\tilde\psi}(t)\rangle
\end{aligned}
\end{equation}
where $\bar H_R(t)=\frac{1}{2}\langle\tilde\psi(t)| H(t)+H^\dagger(t)|\tilde\psi(t)\rangle$, $\bar H_I(t)=\frac{1}{2i}\langle\tilde\psi(t)| H(t)-H^\dagger(t)|\tilde\psi(t)\rangle$. This means that the non-unitary evolution can be divided into two parts: one is the normalized state $|\tilde\psi(t)\rangle$ with a real phase factor $\beta(t)$ which like the state in projective Hilbert space with dynamic and geometric phase, the other part is a pure imaginary phase which corresponds gain and loss introduced by the non-Hermitian part of  $H(t)$. In this sense, the evolution of a state can be described by the evolution of $|\tilde\psi(t)\rangle$ in projective space, and the changing in normalization factor $e^{2\alpha(t)}$ caused by gain or loss and the real phase shift $\beta$ can be directly obtained by an integral on the projective space.

Especially for a state $|\psi_n(t)\rangle$ which is initially a energy eigenstate with energy eigenvalue $E_n(t)$ of $H(t)$ and satisfying the adiabatic condition (\ref{eq:jrjs}),  its evolution can be described by
\begin{equation}
|\psi_n(t)\rangle=e^{\int_0^t \bar{H}_I(t')dt'/\hbar}e^{-i\int_0^t \bar{H}_R(t')dt'/\hbar-\int_0^t\langle\tilde\psi_n(t')|\dot{\tilde\psi_n}(t')\rangle dt'}|\tilde\psi_n(t)\rangle.
\end{equation}
It is interesting to notice that, under the condition (\ref{sncond}) for the self-normalized energy eigenstate in pseudo- and antipseudo- Hermitian system, $|\psi_n\rangle$ has the same form as
\begin{equation}
|\psi_n(t)\rangle=e^{-i\int_0^t E_n(t')dt'/\hbar-\int_0^t\langle E_n(t')|\dot{E_n}(t')\rangle dt'}|E_n(t)\rangle.
\end{equation}
For pseudo-Hermitian system under the condition (\ref{sncond}), the norm $\langle\psi_n(t)|\psi_n(t)\rangle=1$ as $E_n $ is real. While, for the antipseudo-Hermitian system under the condition (\ref{sncond}), the norm $\langle\psi_n(t)|\psi_n(t)\rangle=e^{-2i\int_0^t E_n(t')dt'/\hbar}$ as $E_n $ is pure imaginary. Therefore, the self-normalized energy eigenstates $|\psi_n\rangle$ for antipseudo-Hermitan also require $E_n=0$.

\bibliographystyle{apsrev4-1}
%

\end{document}